\documentclass[twocolumn,aps,showpacs,amsmath,amssymb]{revtex4}
\usepackage{graphicx}
\usepackage{bm}

\begin{document}

\title{Logarithmic Relaxation in Glass-Forming Systems}

\author{W. G\"otze}\author{M. Sperl}
\affiliation{Physik-Department, Technische Universit\"at M\"unchen, 
D-85747 Garching} 

\date{\today}

\begin{abstract}
Within the mode-coupling theory for ideal glass transitions, an analysis for 
the correlation functions of glass-forming systems for states near higher-order
glass-transition singularities is presented. It is shown that the solutions of 
the equations of motion can be asymptotically expanded in polynomials of the 
logarithm of time $t$. In leading order, an $\ln(t)$-law is obtained, and the 
leading corrections are given by a fourth-order polynomial. The correlators 
interpolate between three scenarios. First, there are planes in parameter space
where the dominant corrections to the $\ln(t)$-law vanish, so that the 
logarithmic decay governs the structural relaxation process. Second, the 
dynamics due to the higher-order singularity can describe the initial and 
intermediate part of the $\alpha$-process thereby reducing the range of 
validity of von~Schweidler's law and leading to strong $\alpha$-relaxation 
stretching. Third, the $\ln(t)$-law can replace the critical decay law of the 
$\beta$-process leading to a particularly large crossover interval between the
end of the transient and the beginning of the $\alpha$-process. This may lead 
to susceptibility spectra below the band of microscopic excitations exhibiting
two peaks. Typical results of the theory are demonstrated for models dealing 
with one and two correlation functions.
\end{abstract}

\maketitle

\section{\label{sec:introduction}Introduction}

Within the mode-coupling theory for ideal glass transitions (MCT), the dynamics
of an amorphous system of strongly interacting spherical particles is described
by $M$ functions of time $t$, $\phi_q(t),\;q=1,2,\dots, M$. These are 
autocorrelation functions of density fluctuations with wave-vector modulus $q$ 
chosen from a grid of $M$ values. The theory is based on a closed set of 
coupled nonlinear equations of motion for the $\phi_q(t)$. The coupling 
coefficients in these equations are given in terms of the equilibrium structure
functions. The latter are assumed to be known smooth functions of the control 
parameters of the system like, e.g., the packing fraction $\varphi$ 
\cite{Bengtzelius84}. The MCT equations exhibit fold bifurcations 
\cite{Arnold86} at certain critical values of the control parameters, say at 
$\varphi=\varphi_c$, describing a transition from ergodic liquid dynamics for 
$\varphi<\varphi_c$ to non-ergodic glass dynamics for $\varphi\geqslant
\varphi_c$. The transition is accompanied by the evolution of a slow stretched 
dynamics that was suggested as the explanation of structural relaxation 
observed in glass-forming liquids. The leading-order asymptotic solutions of 
the equations for parameters approaching the transition provide predictions for
the universal properties of glassy dynamics \cite{Goetze92}. These predictions
have been tested extensively against experimental data and 
molecular-dynamics-simulation results \cite{Goetze99,Kob99}. The outcome of 
these tests qualifies MCT as a candidate for a theory of structural relaxation 
in glass-forming systems.

It was shown that schematic MCT models exhibit also higher-order-bifurcation 
singularities like the cusp and swallow-tail bifurcation. The accompanying 
dynamics is utterly different from the one for the fold bifurcation. For 
example, in certain parameter regions, the leading order result reads 
$(\phi(t)-f)\propto-\ln(t/\tau)$ \cite{Goetze88b}. This logarithmic decay is 
equivalent to a susceptibility spectrum that is independent of frequency 
$\omega$, $\chi''(\omega)\propto\omega^0$, or to a $1/f$-noise fluctuation 
spectrum. There are corrections to this leading-order result which alter 
qualitatively the straight $\phi(t)$-versus-$\log t$ lines or the plateaus of 
the $\chi''(\omega)$-versus-$\log(\omega)$ plots. One needs to understand these
corrections if one intends to get an overview of the relaxation scenarios for 
parameters near the higher-order-bifurcation points. It is he goal of this 
paper to provide such understanding by construction of a general theory for the
logarithmic-relaxation law and its leading corrections.

For parameters at a cusp singularity of schematic $M=1$ models, the 
leading-order long-time decay follows the law $(\phi(t)-f)\propto\,1/\ln^2t$. 
This law has been embedded in a leading-order description of the dynamics near 
the singularities in terms of multi-parameter scaling laws \cite{Goetze89d}.
It was shown by L.~Sj\"ogren that dielectric-loss spectra for certain polymers
could be interpreted by this scaling-law description \cite{Sjoegren91}, and 
further work extended his analysis \cite{Flach92,Halalay96,Eliasson01}.
However, it was also demonstrated that the cited decay laws at the critical
points have to be complemented by their leading corrections in order to 
describe the numerical solutions of MCT equations within a time regime 
relevant for data analysis \cite{Goetze89d,Sperl00}. But, so far it was not
possible to evaluate the corrections for the mentioned scaling laws. The 
results of this paper will be obtained along a different route of asymptotic 
expansion of the MCT solutions than followed in Ref.~\cite{Goetze89d}. 

Logarithmic decay of correlations for glassy systems has been observed, for
example, in Monte-Carlo-simulation results for a spin-glass model 
\cite{Binder76}, for photon-correlation data from a dense colloidal suspension
\cite{Bartsch92}, and for optical Kerr-effect data for a van~der~Waals liquid
\cite{Hinze00}. But the present work is motivated by three recent discoveries. 
First, density correlators $\phi_q(t)$ measured by photon-correlation
spectroscopy for colloids of micellar particles demonstrated logarithmic decay
within time windows of two orders of magnitude in size \cite{Mallamace00}.
Second, the MCT equations for a system whose structure was described by 
Baxter's model for sticky hard spheres exhibit cusp bifurcations 
\cite{Fabbian99,Bergenholtz99}. These findings have been corroborated by a 
comprehensive analysis of the glass transitions of a square-well system 
\cite{Dawson01}. Third, logarithmic decay extending over three decades in time
was found in a molecular-dynamics simulation for a system with an interaction 
given by a strong repulsion complemented by a short-ranged attraction 
\cite{Puertas02}. One concludes that higher-order bifurcation singularities are
not restricted to schematic models and that there are reasons to suggest the 
search for such singularities in colloids with short-ranged attraction. It is 
the aim of this paper to provide a detailed discussion of the qualitative 
features, which are characteristic for the relaxation in systems near 
higher-order bifurcations. A set of general formulas shall be derived, which 
could be used as basis for a quantitative analysis of future experiments and 
simulation studies.

The paper is organized as follows. In Sec.~\ref{subsec:basic_mcteq}, the known
general MCT equations for structural relaxation are formulated. Then
(Sec.~\ref{subsec:basic_sing}), these equations are rewritten in a form which
is suited as a basis for an asymptotic solution near bifurcation singularities.
Section~\ref{sec:one} presents the theory for the logarithmic relaxation for 
MCT models dealing with a single correlator and in Sec.~\ref{sec:one_F13} 
quantitative results are discussed for a cusp singularity in an $M=1$ model. 
Section~\ref{sec:q} presents the theory for the general case and in 
Sec.~\ref{sec:BK} further results are discussed for relaxation near a 
swallow-tail singularity for an $M=2$ model. Section~\ref{sec:conclusion} 
summarizes the findings.

\section{\label{sec:basic}Basic Equations}

\subsection{\label{subsec:basic_mcteq}The equations for structural relaxation}

MCT is based on two sets of equations. The first one consists of the exact 
equations of motion for the $M$ density correlators $\phi_q (t),\, q = 1, 2,
\ldots, M$, derived within the Zwanzig-Mori formalism
\begin{subequations}\label{eq:mct}
\begin{equation}\label{eq:mct_eom}
\partial_t^2 \phi_q (t) + \Omega_q^2 \phi_q (t) +
\int_0^t M_q (t - t^\prime) \partial_{t^\prime} \phi_q
(t^\prime)dt^\prime = 0 \,\, .
\end{equation}
The initial conditions read $\phi_q (t=0) = 1,\,\partial_t \phi_q (t=0) = 0$. 
The positive $\Omega_q$ are characteristic frequencies and the kernels $M_q(t)$
are fluctuating-force correlators \cite{Hansen86}. In colloidal suspensions, 
there are contributions to the force due to interactions of a colloid particle
with the solvent particles. These fluctuate on a time scale much shorter than 
the one relevant for the motion of the mesoscopic colloid particles. 
Therefore, one can approximate the corresponding contributions
to the kernel by a white-noise term, and this leads to a friction force 
$\nu_q\partial_t \phi_q(t),\,\nu_q>0$. Compared to this term, one can neglect
the inertia term $\partial_t^2 \phi_q (t)$. One arrives at the analogue of 
Eq.~(\ref{eq:mct_eom}) for colloids, i.e. at an equation of motion where the
underlying dynamics is Brownian rather than Newtonian:
\begin{equation}\label{eq:mct_eomBrown}
\nu_q\partial_t \phi_q(t) + \Omega_q^2 \phi_q (t) +
\int_0^t M_q (t - t^\prime) \partial_{t^\prime} \phi_q
(t^\prime)dt^\prime = 0 \,\, .
\end{equation}
\end{subequations}
The initial conditions are $\phi_q (t=0)=1$.
The kernels are split into regular ones, $M_q^{\rm reg} (t)$, and so-called
mode-coupling kernels $m_q (t)$ describing the cage effect:
$M_q (t)= M_q^{\rm reg} (t) + \Omega_q^2 m_q (t)$. The regular terms describe 
normal liquid effects like binary collisions in conventional liquids or
hydrodynamic interactions in colloids. The crucial step in the derivation of 
MCT is the application of Kawasaki's factorization approximation in order to 
express $m_q (t)$ as functionals ${\cal F}_q$ of the $M$ correlators 
$\phi_k(t)$
called mode-coupling functionals. For simple systems, they
are quadratic polynomials, whose coefficients are given in terms of the 
equilibrium structure functions \cite{Bengtzelius84,Goetze91b}. 
The functionals depend smoothly on, say, $N$ control parameters to be combined
to a control parameter vector $V=(V_1,\ldots,V_N)$. In conventional liquids, 
the packing fraction and the temperature may be the control parameters. 
In colloids, one of the control parameters may be the attraction strength, 
which can be changed by modifying the solvent. Let us note the functionals
by ${\cal F}_q[V,\tilde{f}_k]$. For 
$0\leqslant |\tilde{f}_k  |\leqslant 1,\,k= 1, 2,\ldots, M$, they
can be written as Taylor series with non-negative coefficients.
Thus, the second set of MCT equations is 
\begin{equation}\label{eq:mct_kernel}
m_q (t) = {\cal F}_q \left[V, \phi_k (t) \right] \,\, .
\end{equation}
Specifying the functionals  ${\cal F}_q$, the regular kernels 
$M_q^{\rm reg} (t)$, and the frequencies $\Omega_q$ or $\nu_q$,
Eqs.~(\ref{eq:mct_eom}) or Eqs.~(\ref{eq:mct_eomBrown})
together with Eq.~(\ref{eq:mct_kernel}) are closed. 
In the present paper, a topologically stable singularity of Eqs.~(\ref{eq:mct})
and (\ref{eq:mct_kernel}) shall be discussed in full generality. Therefore,
microscopic details are not of concern. 

It will be convenient to discuss dynamics in the domain of complex
frequencies $z, \, {\rm Im}\, z > 0$. This can be achieved by Laplace
transformation of functions of time, say $F (t)$, to functions of
$z$ denoted by ${\cal L} [F (t)] (z) = i \int_0^\infty \exp (izt)
F(t) dt$. Equations (\ref{eq:mct_eom}) and (\ref{eq:mct_kernel}) together 
with the initial conditions are equivalent to a fraction
representation of $\phi_q (z) = {\cal L} [\phi_q (t)] (z)$ in
terms of $M_q^{\rm reg} (z) = {\cal L} [M_q^{\rm reg} (t)] (z)$
and $m_q (z) = {\cal L} [m_q (t)] (z)$,
\begin{subequations}\label{eq:mct_laplace}
\begin{equation}\label{eq:mct_laplace_Newton}
\phi_q (z) = - 1 / \{ z -  \Omega_q^2 / \left[ z + M_q^{\rm reg}
(z) + \Omega_q^2 m_q (z) \right] \} \,\, .
\end{equation}
The analogue for the colloid dynamics is derived from 
Eq.~(\ref{eq:mct_eomBrown}):
\begin{equation}\label{eq:mct_laplace_Brown}
\phi_q (z) = - 1 / \{ z -  \Omega_q^2 / \left[i\nu_q + M_q^{\rm reg}
(z) + \Omega_q^2 m_q (z) \right] \} \,\, .
\end{equation}
\end{subequations}

There are two possibilities for the solutions of the preceding
equations. Either, all long-time limits of the correlators vanish
as expected for an ergodic system. States with such control
parameters $V$ are referred to as liquids. Or, there
may be nonvanishing long-time limits $f_q = \phi_q (t \to
\infty), \, 0 < f_q \leqslant 1$, as expected for non-ergodic
systems. States with such control parameters $V$ are referred to
as glasses, and $f_q$ is called the glass-form factor.
 Changing $V$, it may happen that there are values
$V^c$ where one changes from a liquid to a glass --- these are
the ideal liquid-glass transitions discussed within MCT \cite{Bengtzelius84}.
For glass states, $\phi_q (z)$ exhibits a zero-frequency pole,
$\phi_q (z \to 0) \sim - f_q /z$. Because of Eq.~(\ref{eq:mct_kernel}), 
a similar
statement holds for the kernel, $m_q (z \to 0) \sim - {\cal F}_q
[V, f_k] / z$. Hence, for frequencies tending to zero, kernel $m_q
(z)$ becomes arbitrarily large compared to the term $z + M_q^{\rm
reg} (z)$ or the term $i\nu_q + M_q^{\rm reg} (z)$, respectively. 
Because of continuity, also for states with control
parameters near the ones for glass states $\Omega_q^2 m_q (z)$ is
very large compared to $z + M_q^{\rm reg} (z)$ or to 
$i\nu_q + M_q^{\rm reg} (z)$. Under the
specified conditions, Eqs.~(\ref{eq:mct_laplace_Newton}, b) 
 simplify to \cite{Goetze92}
\begin{equation}\label{eq:mct_laplace_general}
\phi_q (z) = - 1 / [z- 1 / m_q (z)] \,\, .
\end{equation}
This equation exhibits most clearly the fraction representation of correlators,
which is the essence of the Zwanzig-Mori theory. It shows that there is a 
one-to-one correspondence between density fluctuations and force fluctuations.
There is no separation between the time scales for the particle motion within
cages and for the particles forming the cage. Therefore, the correlators 
$\phi_q(t)$ and the kernels $m_q(t)$ have to be calculated self-consistently
--- this is the essence of MCT. Equation~(\ref{eq:mct_laplace_general}) is 
scale invariant. With $\phi_q(t)$, also $\phi^x_q(t)=\phi_q(x\cdot t)$ is 
a solution for any $x>0$. The scale for the high frequency dynamics is 
determined by the transient motion and this is governed by $M_q^{\rm
reg} (t)$ and $\Omega_q$ or $\nu_q$. But these quantities do not occur anywhere
in Eq.~(\ref{eq:mct_laplace_general}). Thus, Eq.~(\ref{eq:mct_laplace_general})
can fix the solution only up to some time scale.

Equations~(\ref{eq:mct_kernel}) and (\ref{eq:mct_laplace_general}) are the
MCT equations for structural relaxation. In particular, they are the basis of 
the asymptotic expansions for the long-time dynamics for control parameters 
near bifurcation points.

\subsection{\label{subsec:basic_sing}The equations for structural relaxation
near glass-transition singularities}

In this section, the concept of a glass-transition singularity shall be 
reviewed. The equations of motion shall be rewritten in a form where the 
small quantities, which characterize the relaxation near such a singularity,
appear transparently.

To simplify the notation of the following equations, the Laplace transform 
shall be modified by a factor of $(-z)$:
\begin{eqnarray}\label{eq:strafo}
{\cal S}  \left[F(t)\right] (z) & =&
(-iz) \int_0^\infty\exp(izt)\, F(t)\, dt \,\, .
\end{eqnarray}
This linear mapping of functions of time to functions of frequency leaves 
constants invariant: $F(t)=c$ implies ${\cal S}[c](z)=c$. 
Let $F(t) = \langle X^*(t)X\rangle /(k_{\rm B}T)$ denote a correlation 
function for variable $X$ determined for temperature $T$. Then, the dynamical 
susceptibility for frequency $\omega$ can be written as $\chi(\omega) = 
F(t=0) - {\cal S}\left[F(t)\right] (\omega+i0)$ \cite{Hansen86}. Thus,
${\cal S}  \left[F(t)\right] (z)$ denotes the non-trivial part of a dynamical
susceptibility. 

Equation~(\ref{eq:mct_laplace_general}) can be rewritten as
\begin{equation}\label{eq:mct_strafo}
{\cal S}  \left[\phi_q (t) \right] (z)  / \{ 1 - {\cal S} \left[
\phi_q (t) \right] (z) \} = {\cal S} \left[ {\cal F}_q \left[ V,
\phi_k (t) \right] \right] (z) \,\, .
\end{equation}
Equation $\phi_q(t\rightarrow\infty)=f_q$ is equivalent to
${\cal S}\left[\phi_q(t)\right](z\rightarrow 0)=f_q$. Similarly, one obtains
${\cal S}\left[{\cal F}_q[V,\phi_k(t)]\right](z\rightarrow 0)=
{\cal F}_q[V,f_k]$. The $z\rightarrow 0$ limit of 
Eq.~(\ref{eq:mct_strafo}) yields a set of $M$ implicit equations for the $M$ 
glass-form factors $f_q$ \cite{Bengtzelius84}:
\begin{equation}\label{eq:singularity}
f_q / (1 - f_q) = {\cal F}_q \left[V, f_k \right] \,\, .
\end{equation}
This equation may have other solutions, say $\tilde{f}_q$. The glass-form
factor is distinguished by the maximum theorem: $\tilde{f}_q\leqslant f_q,\,
q=1,\,\dots,\,M$ \cite{Goetze91b}.

Let $V^c$ denote some reference state. The long-time limits of the 
correlators for this state shall be denoted by $f^c_q$. The correlators shall 
be written in terms of new functions $\hat{\phi}_q(t)$
\begin{equation}\label{eq:hat_phi}
\phi_q (t) = f_q^c + (1 - f_q^c) \hat \phi_q (t) \,\, .
\end{equation}
The functional ${\cal F}_q[V,\phi_k(t)]$ can be rewritten
as a Taylor series in $\hat{\phi}_k(t)$, using the coefficients
\begin{subequations}\label{eq:Aqk}
\begin{eqnarray}\label{eq:Aqk_def}
A_{q k_1 \cdots k_n}^{(n)} (V) & = & \frac{1}{n!} (1 - f_q^c)
\{\partial^n {\cal F}_q \left [V, f_k^c \right ] / \partial
f_{k_1}^c \cdots
\partial f_{k_n}^c \} \nonumber\\
& \times &(1 - f_{k_1}^c) \cdots (1 - f_{k_n}^c)   \,\, .
\end{eqnarray}
These shall be split in the values for the reference state, 
$A_{q k_1 \cdots k_n}^{(n) c} = A_{q k_1\cdots k_n}^{(n)} (V^c)$, 
and remainders:
\begin{equation}\label{eq:deltaA_def}
A_{q k_1 \cdots k_n}^{(n) } (V) = A_{q k_1 \cdots k_n}^{(n) c} +
\hat{A}_{q k_1 \cdots k_n}^{(n)} (V) \,\, .
\end{equation}
Let us consider a path in control-parameter space given by
$V (\epsilon) = (V_1 (\epsilon), \ldots V_N (\epsilon))$. The $N$
components of $V(\epsilon)$ are smooth functions of the path
parameter $\epsilon$, and the tangent vector $d V (\epsilon) / d
\epsilon$ must not vanish. Let us choose $V (\epsilon = 0) =
V^c$, so that $\epsilon$ can be considered as a distance parameter
specifying the neighborhood of $V^c$. One gets $V (\epsilon) = T
\epsilon +{\cal O}(\epsilon^2)$, with $T = d V (0) / d \epsilon$ being
the tangent vector of the path at $V^c$. The mode-coupling
functional is a smooth function of $V$, i.e.
\begin{equation}\label{eq:deltaA_order}
\hat{A}_{q k_1\dots k_n}^{(n)} (V) ={\cal O}(\epsilon) \,\, .
\end{equation}
\end{subequations}
The details of the $\epsilon$-dependence of the coefficients are
not important. The parameter $\epsilon$ is introduced mainly as a
means for bookkeeping in the following expansions in $V - V^c$.
Expanding the left hand side of Eq.~(\ref{eq:mct_strafo}) in powers of 
${\cal S}\left[\hat{\phi}_q\right](z)$, 
one can rewrite this equation in the form
\begin{subequations}\label{eq:J}
\begin{equation}\label{eq:J_def}
\left[ \delta_{q k} - A_{q k}^{(1) c} \right] {\cal S} [ \hat
\phi_k (t) ] (z) = J_q (z) \,\, ,
\end{equation}
\begin{eqnarray}\label{eq:J_exp}
J_q (z) & = & \hat{A}_q^{(0)} (V) + \hat{A}_{q k}^{(1)} (V)
{\cal S} [ \hat \phi_k (t) ] (z) \nonumber \\  & + &
\sum_{n = 2}^\infty \{ A_{q k_1 \cdots k_n}^{(n)} (V) {\cal S}
[ \hat \phi_{k_1} (t) \cdots  \hat \phi_{k_n} (t) ] (z)\nonumber\\
&&\qquad- {\cal S} [ \hat \phi_q (t)]^n (z) \} \,\, .
\end{eqnarray}
\end{subequations}
Here and in the following, summation over pairs of equal labels $k$ is implied.
These equations are equivalent to Eqs.~(\ref{eq:mct_kernel}) and 
(\ref{eq:mct_laplace_general}) for the structural relaxation.
The small quantities to be used for the asymptotic solution are the 
coefficients $\hat{A}^{(n)}_{qk_1\dots k_n}$ and the functions 
$\hat{\phi}_q(t)$ or ${\cal S} \left[ \hat \phi_q (t) \right] (z)$,
respectively.

Specializing Eqs.~(\ref{eq:J}) to the $z\rightarrow 0$ limit, one gets the
equation for $\hat{f}_q=\hat{\phi}_q(t\rightarrow\infty)$:
\begin{eqnarray}\label{eq:matrix}
\bigl [ \delta_{q k} & - & A_{q k}^{(1) c} \bigr ] \hat f_k =
\hat{A}_q^{(0)} (V) + \hat{A}_{q k}^{(1)} (V) \hat f_k \nonumber \\
 & + &  \sum_{n = 2 }^\infty \left [ A_{q k_1 \cdots k_n}^{(n)}
(V) \hat f_{k_1} \cdots \hat f_{k_n} - \hat{f}_q^n \right ] \,\, .
\end{eqnarray}
This is a rewriting of Eq.~(\ref{eq:singularity}) so that small deviations of 
$f_q$ from $f^c_q$ and $V$ from $V^c$ are explicit. The $M$ by $M$ matrix 
$\bigl [ \delta_{q k} - A_{q k}^{(1) c} \bigr ]$ is the Jacobian of the set
of implicit equations~(\ref{eq:singularity}) for the reference solution $f^c_q$
at $V=V^c$. This Jacobian consists of a unit matrix $\delta_{q k}$
and a matrix $A_{q k}^{(1) c}$ of positive elements. The Frobenius
theorems imply that, generically, this matrix has a non degenerate maximum
eigenvalue $E^c > 0$. All other eigenvalues have a modulus smaller
than $E^c$ \cite{Gantmacher74b}. It is a subtle property of MCT
that $E^c \leqslant 1$ \cite{Goetze95b} . If $E^c < 1$, the
implicit-function theorem guarantees that all states $V$ for sufficiently 
small $\epsilon$ are states whose 
long-time limits $f_q=f^c_q+(1-f^c_q)\hat{f}_q$
depend smoothly on $\epsilon$.
In the following, the reference state $V^c$ shall be specialized so that
\begin{equation}\label{eq:eigenvalue}
E^c = 1 \,\, .
\end{equation}
In this case, $V^c$ is a bifurcation point of Eq.~(\ref{eq:singularity}). The 
$f_q$ are singular functions of $\epsilon$ for $\epsilon \to 0$, and therefore
$V^c$ is referred to as a glass-transition singularity. Since $E^c$ is 
non-degenerate, the possible bifurcations are from the so-called cuspoid family
$A_l, \, l = 2,3, \ldots$. The bifurcation singularity $A_l$ is topologically 
equivalent to that for the zeroes of a real polynomial of degree $l$ 
\cite{Arnold86} . The $A_2$, also called fold bifurcation, is the generic 
singularity obtained by varying a single control parameter. The liquid-glass
transition of MCT is of this type. In this paper, the dynamics near a higher 
order singularity $A_l,\,l\geqslant3$, shall be analyzed.

\section{\label{sec:one}Relaxation Described by One-Component Models}

It will be shown in Sec.~\ref{sec:q} that each iteration step of the 
asymptotic solution of the equations of motion splits into two parts. 
First, one has to reduce the problem of calculating $M$ correlators
to the one of calculating the projection of the correlators on the
dangerous eigenvector of the above defined Jacobian. 
Second, one has to solve the equation for the 
projection. In this section, the second problem will be studied, which is 
equivalent to a discussion of $M=1$ models.

\subsection{\label{subsec:classification}Classification of glass-transition
singularities}

One component models deal with a single correlator $\phi(t)$, a single 
glass-form factor $f$, etc. All matrix indices can be dropped in the formulas 
of Sec.~\ref{sec:basic}. The one-by-one matrix $A^{(1)c}_{qk}$ is identical 
with its maximum eigenvalue $E^c$. Because of Eq.~(\ref{eq:eigenvalue}), the 
left-hand side of Eq.~(\ref{eq:matrix}) vanishes. The equation for 
$\hat{f}$ reads $\epsilon_1(V)+\epsilon_2(V)\hat{f}+
\sum_{n\geqslant 2}[\hat{A}^{(n)}(V)-\mu_n]\hat{f}^n=0$, where the 
abbreviations are used
\begin{equation}\label{eq:mu_def}
\mu_n  = 1 - A^{(n)c}\,,\quad
\epsilon_n(V) = \hat{A}^{(n-1)}(V)\,,\quad n=1,2,\dots
\,\, .
\end{equation}
The singularity exhibited by $\hat{f}$ for $\epsilon$ tending to zero depends
on the number of successive vanishing coefficients 
$\mu_n$. A singularity of index $l,\,l\geqslant 2$, 
shall be defined by
\begin{equation}\label{eq:A_l:hierarchy}
\mu_1=\mu_2=\dots=\mu_{l-1}=0\;,\quad\mu_l\neq 0\,\, .
\end{equation}
The equation for $\hat{f}$ reads
\begin{equation}\label{eq:A_l_mu_l}\begin{split}
\mu_l\hat{f}^l =&\epsilon_{l-1}(V)\hat{f}^{l-2}
			+\epsilon_{l-2}(V)\hat{f}^{l-3}+\dots+\epsilon_1(V)\\
+	&\{\epsilon_l(V)\hat{f}^{l-1}
		+ \epsilon_{l+1}(V)\hat{f}^l
	+ \sum_{n\geqslant l+1}\left[A^{(n)}(V)-1\right]\hat{f}^n
\} \,\,.
\end{split}\end{equation}
The implicit-function theorem can be used to show that there is a smooth 
invertible transformation of the $l$ variables $\left(\epsilon_1, \epsilon_2,
\ldots, \epsilon_{l-1}, \hat{f}\right)$ which eliminates the curly bracket 
in Eq.~(\ref{eq:A_l_mu_l}). Thus, the singularities described by this equation
are topologically equivalent to the ones described by the first line, i.e. by
the zeros of a polynomial of degree $l$. In Arnol'd's terminology 
\cite{Arnold86}, such singularity is referred to as $A_l$. Because of 
Eq.~(\ref{eq:deltaA_order}), the $\epsilon_n(V)$ are of 
order $\epsilon$ and shall be referred to as separation parameters.

The simplest glass-transition singularity is the $A_2$.
In this case,  there is only one relevant control parameter $\epsilon_1(V)$.
One infers from Eq.~(\ref{eq:A_l_mu_l}) that there is a discontinuous change of
$\hat{f}$ at the surface specified by $\epsilon_1(V)=0$. The bifurcation 
dynamics is characterized  by power-law decay and there appear power-law 
dependencies of the relaxation scales on $|\epsilon_1(V)|$. All exponents in 
these laws are to be calculated from $\lambda=1-\mu_2$, which is called the 
exponent parameter \cite{Goetze91b}. The transition surface has a boundary 
that is determined by $\lambda=1$, i.e. by $\mu_2=0$. These endpoints are the 
higher-order singularities. The $A_3$ and $A_4$ are also referred to as cusp
and swallow-tail singularities, respectively.

\subsection{\label{subsec:one_asy}Equations for an asymptotic solution}

Let us specialize Eqs.~(\ref{eq:J_def}) and (\ref{eq:J_exp}) for $M=1$.
Let us also express the coefficients $A^{(n)}(V)$ in terms of $\mu_n$
and $\epsilon_n(V)$:
\begin{equation}\label{eq:problem}
\begin{array}{cclcl}
0 = \epsilon_1(V) &+& (1-\mu_2){\cal S}[\hat{\phi}^2(t)](z) 
			& - &{\cal S}[\hat{\phi}(t)]^2(z) \\
+ \epsilon_2(V) {\cal S}[\hat{\phi}(t)](z)
		&+& (1-\mu_3){\cal S}[\hat{\phi}^3(t)](z) 
			& - &{\cal S}[\hat{\phi}(t)]^3(z) \\
+ \epsilon_3(V) {\cal S}[\hat{\phi}^2(t)](z)
		&+& (1-\mu_4){\cal S}[\hat{\phi}^4(t)](z) 
			& - &{\cal S}[\hat{\phi}(t)]^4(z)\\
 +\; \cdots \,\, .&&&&
\end{array}
\end{equation}
This suggests an expansion of the 
solution in powers of $|\epsilon|^{1/2}$. With
$G^{(n)} (t) = {\cal O} (|\epsilon|^{n/2})$, let us write
\begin{equation}\label{eq:solution_expansion}
\hat{\phi}(t) = G^{(1)} (t) + G^{(2)} (t) + G^{(3)} (t) + \cdots \,\, .
\end{equation}
The first line of Eq.~(\ref{eq:problem}) is of order $|\epsilon|$ and it 
provides a nonlinear integral equation for $G^{(1)} (t)$. 
The contributions to this line which are of order $|\epsilon|^{3/2}$ together
with the leading terms of the second line provide a linear integral equation
for $G^{(2)} (t)$, etc. This procedure will yield the desired 
asymptotic expansion provided the indicated integral equations define 
meaningful solutions. This is indeed the case as shall be demonstrated below 
by explicit construction of the $G^{(n)} (t)$. To proceed, the following 
discussion shall be restricted to higher-order singularities by requiring
\begin{equation}\label{eq:lambda_unity}
\mu_2 = 0\,\, .
\end{equation}

\subsection{\label{subsec:one_lead}The leading contribution}

The equation for the leading contribution to the correlator at a glass 
transition $A_l$ with $l\geqslant 3$ reads
\begin{equation}\label{eq:solution_lead}
\epsilon_1 (V) + {\cal S} [{G^{(1)}}^2 (t)](z)-{\cal S}[G^{(1)}(t)]^2(z)= 0 
\, \, .
\end{equation}
The formulas for the Laplace transforms of the logarithm and its
square imply ${\cal S} [\ln (t)] (z) = \ln (i/z) - \gamma$
and ${\cal S} [\ln^2 (t)] (z) = \ln^2 (i/z) - 2\gamma \ln
(i/z) + \gamma^2 + (\pi^2 / 6)$, where $\gamma = 0.577 \ldots$ is
Euler's constant (cf. Appendix~\ref{sec:laplace}, Eq.~(\ref{eq:Strafo_log})). 
Hence, Eq.~(\ref{eq:solution_lead}) is solved by 
$- B \ln (t)$ if $\epsilon_1 (V) + (B^2 \pi^2 /6) = 0$. 
Since the correlators are monotonically decreasing functions of $t$ 
\cite{Goetze95b}, one must require $B>0$. One concludes that a solution 
is given by 
\begin{equation}\label{eq:G1}
G^{(1)} (t) = - B \ln (t) \,\, , \quad B = \sqrt{\left[ - 6
\epsilon_1 (V) / \pi^2 \right]} \, \, ,
\end{equation}
provided the control parameters $V$ obey
\begin{equation}\label{eq:eps1_negativ}
\epsilon_1 (V) < 0 \, \, .
\end{equation}
A more general solution is $\tilde{G}^{(1)} (t) = G^{(1)} (t) + c$, where
$c$ can be any real constant. Introducing $x = \exp (- c / B)$,
one gets $\tilde{G}^{(1)} (t) = G^{(1)} (xt)$. Thus, the generalization is the
one implied by the scale invariance of the basic 
Eq.~(\ref{eq:mct_laplace_general}). It shall not be considered here. 
Rather it shall 
be accounted for at the end of all calculations by rescaling $t$ to $t / \tau$.
Ignoring corrections of order $|\epsilon |$, one derives from 
Eqs.~(\ref{eq:hat_phi}),
(\ref{eq:solution_expansion}), 
and (\ref{eq:G1}) the leading approximation for the correlator 
\cite{Goetze88b}:
\begin{equation}\label{eq:phi_one_lead}
\phi(t) = f^c -(1-f^c) B \ln (t/\tau)\,\, .
\end{equation}

Let us anticipate that the smooth function $\epsilon_1 (V)$
is generic for $V$ near $V^c$ and has a nonvanishing
gradient. Then, $\epsilon_1 (V) = 0$ defines a smooth
surface through $V^c$ in the control-parameter
space. It separates the neighborhood of the glass-transition
singularity $V^c$ into a strong-coupling side where $\epsilon_1 (V)> 0$ 
and a weak-coupling side where $\epsilon_1 (V) < 0$. The
results of this paper refer to the latter regime.

\subsection{\label{subsec:one_corr1}The leading correction}

In order to solve Eq.~(\ref{eq:problem}) up to order $|\epsilon|^{3/2}$,
one has to incorporate from the first line the contribution 
$2 {\cal S}[G^{(1)}(t) G^{(2)}(t)](z) -
  2 {\cal S}[G^{(1)}](z)\, {\cal S}[G^{(2)}(t)](z)$, one has to evaluate 
the second line with $\hat{\phi}$ replaced by $G^{(1)}(t)$, and one can ignore
all other terms. Hence, the equation for the leading correction $G^{(2)}(t)$
can be written in the form
\begin{equation}\label{eq:problem_corr}
{\cal T} \left[ G^{(2)} (t) \right] (z) = f^{(2)} (z) \,\, .
\end{equation}
Here, the linear integral operator ${\cal T}$ is defined by
\begin{equation}\label{eq:Ttrafo}
{\cal T} \left[ G (t) \right] (z) = {\cal S} \left[ \ln (t) G
(t) \right] (z) - {\cal S} \left[ \ln (t)  \right] (z) {\cal S}
\left[ G (t)  \right] (z)\, \, ,
\end{equation}
and the inhomogeneity of Eq.~(\ref{eq:problem_corr}) reads
\begin{equation}\label{eq:f2}
\begin{split}f^{(2)} (z) = - \left\{\epsilon_2 (V) {\cal S} [G^{(1)}  (t) ] (z)
- \mu_3 {\cal S}  [{G^{(1)}}^3  (t) ] (z) \right.\\ \left.
+ 2 \zeta \{ {\cal S} [{G^{(1)}}^3  (t)] (z) -
{\cal S} [G^{(1)}(t)]^3 (z) \}\right\}/(2B)\,\, .
\end{split}
\end{equation}
A factor $2 \zeta$ has been introduced for later convenience. For the 
study of $M=1$ models, one has to substitute $\zeta=1/2$.

The solution of Eq.~(\ref{eq:solution_lead}) was built on the equations
${\cal T}[c](z) = 0,\, {\cal T}\left[\ln t\right](z) = \pi^2/6$.
These formulas are generalized in Appendix~\ref{sec:laplace} by constructing 
polynomials $p_n (x)$ of degree $n\geqslant 1$ with the properties:
\begin{subequations}\label{eq:solution_corr1}
\begin{equation}\label{eq:solution_corr1_poly}
p_n (x) = b_{n, 1} x + b_{n, 2} x^2 + \cdots + b_{n, n-1} x^{n -
1} + x^n\,\,,
\end{equation}
\begin{equation}\label{eq:solution_corr1_Ttrafo_log_poly}
{\cal T} \left [ p_n (\ln (t))  \right ] (z) = n(\pi^2 / 6)\,
\ln^{n - 1} (i / z)  \,\, .
\end{equation}
\end{subequations}
These polynomials are a convenient tool to solve the equation
\begin{subequations}\label{eq:Tproperties}
\begin{equation}\label{eq:Tproperties_T}
{\cal T} \left[ g (t) \right] (z) = f (z)
\end{equation}
for inhomogeneities $f (z)$ which are polynomials in $\ln (i/z)$,
 \begin{equation}\label{eq:Tproperties_f}
f (z) = \sum_{n = 0}^{m} a_n \ln^n (i/z) \,\, .
\end{equation}
Obviously,
\begin{equation}\label{eq:Tproperties_g}
g (t) = \sum_{n = 1}^{m + 1} \left[ a_{n-1} / (n \pi^2 / 6)
\right] p_n (\ln (t)) \,\, .
\end{equation}
\end{subequations}

Using Eq.~(\ref{eq:G1}) and applying Eqs.~(\ref{eq:Strafo_log}) 
and~(\ref{eq:Strafo_log_diff}) for the evaluation of the
transformations of the powers of $\ln (t)$, one can write $f^{(2)}(z)$
in the form of Eq.~(\ref{eq:Tproperties_f}) for $m=3$. The coefficients
are linear functions of $\epsilon_1(V)$ and $\epsilon_2(V)$:
\begin{subequations}\label{eq:acoeffs}
\begin{equation}\label{eq:acoeffs_a0}
a_0 = \bigl [ (6 \zeta / \pi^2) (\Gamma_3 - \Gamma_1^3) - (3 \mu_3
/ \pi^2) \Gamma_3 \bigr ] \epsilon_1 (V) - (\Gamma_1 / 2)
\epsilon_2 (V) \,,
\end{equation}
\begin{equation}\label{eq:acoeffs_a1}
a_1 = \bigl [ 3 \zeta - (9 \mu_3 / \pi^2) \Gamma_2 \bigr ]
\epsilon_1 (V) - ( 1 / 2) \epsilon_2 (V) \,\, ,
\end{equation}
\begin{equation}\label{eq:acoeffs_a2}
a_2 =  - (9 \mu_3 / \pi^2) \Gamma_1 \epsilon_1 (V)  \, ,
\quad
a_3 =  - (3 \mu_3 / \pi^2)  \epsilon_1 (V)  \, .
\end{equation}
\end{subequations}
Here, $\Gamma_k=d^k\Gamma(1)/dx^k$ denotes the $k$th derivative of the gamma 
function at unity. One concludes that $G^{(2)} (t) = g(t)$, where 
Eq.~(\ref{eq:Tproperties_g}) is to be used with $m = 3$:
\begin{subequations}\label{eq:Bcoeffs}
\begin{equation}\label{eq:G2}
G^{(2)} (t)  = \sum_{j = 1}^4 B_j \ln^j (t)\,\,.
\end{equation}
The coefficients are derived with Eqs.~(\ref{eq:solution_poly}a--c):
\begin{equation}\label{eq:Bcoeffs_B1}
B_1 =  (0.44425 \zeta - 0.065381 \mu_3)  \epsilon_1 (V) - 0.22213
\epsilon_2 (V) \,\, ,
\end{equation}
\begin{equation}\label{eq:Bcoeffs_B2}
B_2 = (0.91189 \zeta + 0.068713 \mu_3) \epsilon_1 (V) - 0.15198
\epsilon_2 (V) \,\, ,
\end{equation}
\begin{equation}\label{eq:Bcoeffs_B3}
B_3 = - 0.13504 \mu_3 \epsilon_1 (V) \,\, ,
\quad B_4 = - 0.046197 \mu_3 \epsilon_1 (V) \,\, .
\end{equation}
\end{subequations}
Dropping corrections of order $|\epsilon|^{3/2}$, the solution up to 
next-to-leading order reads
\begin{eqnarray}\label{eq:phi_corr1}
\phi (t) - f^c =  (  1 - f^c) \bigl [ (-B + B_1) \ln (t /
\tau) + B_2 \ln^2 (t / \tau) \nonumber &&\\ +  B_3 \ln^3 (t
/ \tau) + B_4 \ln^4 (t / \tau) \bigr ]&&   .
\end{eqnarray}
A singularity $A_l$ with $l\geqslant 4$ implies $\mu_3=0$. In this case,
the formula simplifies because $B_3=B_4=0$. 

The described procedure can be continued. To solve Eq.~(\ref{eq:problem}) up to
order $\epsilon^2$, one derives the analogue to Eq.~(\ref{eq:problem_corr}):
${\cal T}[G^{(3)}(t)](z) = f^{(3)}(z)$. Function $f^{(3)}(z)$ has the form of
Eq.~(\ref{eq:Tproperties_f}) with $m=6$, where the coefficients $a_j$ depend 
on the parameters $\epsilon_1(V),\,\epsilon_2(V),\,\epsilon_3(V),\,\mu_3$ and 
$\mu_4$. As a result, one gets 
\begin{equation}\label{eq:corr2}
G^{(3)}(t) = \sum_{j=1}^7 C_j \ln^j(i/z)
\,\, ,
\end{equation}
where $C_j={\cal O}(|\epsilon|^{3/2})$.

\section{\label{sec:one_F13}Results for a one-component model}

The simplest example for a generic cusp bifurcation is provided by an $M=1$ 
model with the mode-coupling functional 
${\cal F} [V, \tilde f] = v_1 \tilde f + v_3 \tilde f^3$. This model was 
derived originally within a microscopic theory of spin-glass transitions 
\cite{Goetze84b}. It shall be used here in order to demonstrate several 
implications of our theory. Let us use the model with a Brownian microscopic 
dynamics so that Eqs.~(\ref{eq:mct_eomBrown}) and (\ref{eq:mct_kernel}) 
specialize to 
\begin{subequations}\label{eq:F13_def}
\begin{eqnarray}\label{eq:F13_def_eom}
\tau_1\partial_t\phi(t)+\phi(t)+\int_0^tm(t-t')\partial_{t'}\phi(t')\, dt'
	= 0\,\,,\\
\label{eq:F13_def_m}
m(t) = v_1 \phi(t) + v_3 \phi^3(t)
\,\,.
\end{eqnarray}
\end{subequations}
The two coupling constants $v_1\geqslant 0$ and $v_3\geqslant 0$ are considered
as the components of the control-parameter vector $V=(v_1,v_3)$.

Figure~1 reproduces the phase diagram \cite{Goetze88b,Goetze91b}. It is 
obtained from the largest of the solutions for $f^c$ of 
Eq.~(\ref{eq:singularity}), i.e., $v_1^c f^c + v_3^c {f^c}^3 = f^c/(1-f^c)$, 
and Eq.~(\ref{eq:eigenvalue}), i.e., $v_1^c+3 v_3^c  {f^c}^2 = 1/(1-f^c)^2,\,
0\leqslant f^c\leqslant 1$. There are two transition lines. The first one is 
the straight horizontal line of degenerate $A_2$ bifurcations: 
$v_1^c=1,\,0\leqslant 
v_3^c\leqslant 4,\,f^c=0$. Crossing this line by increasing $v_1$,
$f=\phi(t\rightarrow\infty)$ increases continuously. The second one is the 
smooth curve of $A_2$ singularities $V^{(2)c}$ shown as a heavy full line. It 
starts at $v_1^{(2)c}=0,\,v_3^{(2)c}=27/4,\,f^{(2)c}=2/3$. With decreasing 
$v_3^{(2)c}$, $f^{(2)c}$ decreases along the line. For $v_1^{(2)c}=1,\,
v_3^{(2)c}=4$ one gets $f^{(2)c}=1/2$. Decreasing $f^{(2)c}$ further, the line
reaches the endpoint that is marked by a circle. This is the
$A_3$ singularity $V^c$ specified by
\begin{equation}\label{eq:F13_crit}
v_1^c = 9/8 \, , \quad v_3^c = 27/8 \, , \quad f^c = 1/3 \, ,
\quad \mu_3 = 1/3 \,\, .
\end{equation}
The two separation parameters are obtained from Eqs.~(\ref{eq:deltaA_def})
and (\ref{eq:mu_def}) as linear functions of the parameters differences
$\hat{v}_{1,3} = v_{1,3}-v_{1,3}^c$:
\begin{equation}\label{eq:F13_epsilon}
\epsilon_1 (V) = (2/81) [ 9 \hat{v}_1 + \hat{v}_3],\,
\epsilon_2 (V) = (4/27) [ 3 \hat{v}_1 + \hat{v}_3] \,\, .
\end{equation}
These formulas determine the coefficient $B$ in Eq.~(\ref{eq:G1}) and 
$B_1$--$B_4$ in Eqs.~(\ref{eq:Bcoeffs}). The scales $\tau$ for the results in 
Eqs.~(\ref{eq:phi_one_lead}),~(\ref{eq:phi_corr1}), and~(\ref{eq:corr2})
are determined as the time where the correlator crosses the critical
form factor: $\phi(\tau)=f^c$.

\begin{figure}[Hhtb]
\noindent\includegraphics[width=0.42\textwidth]{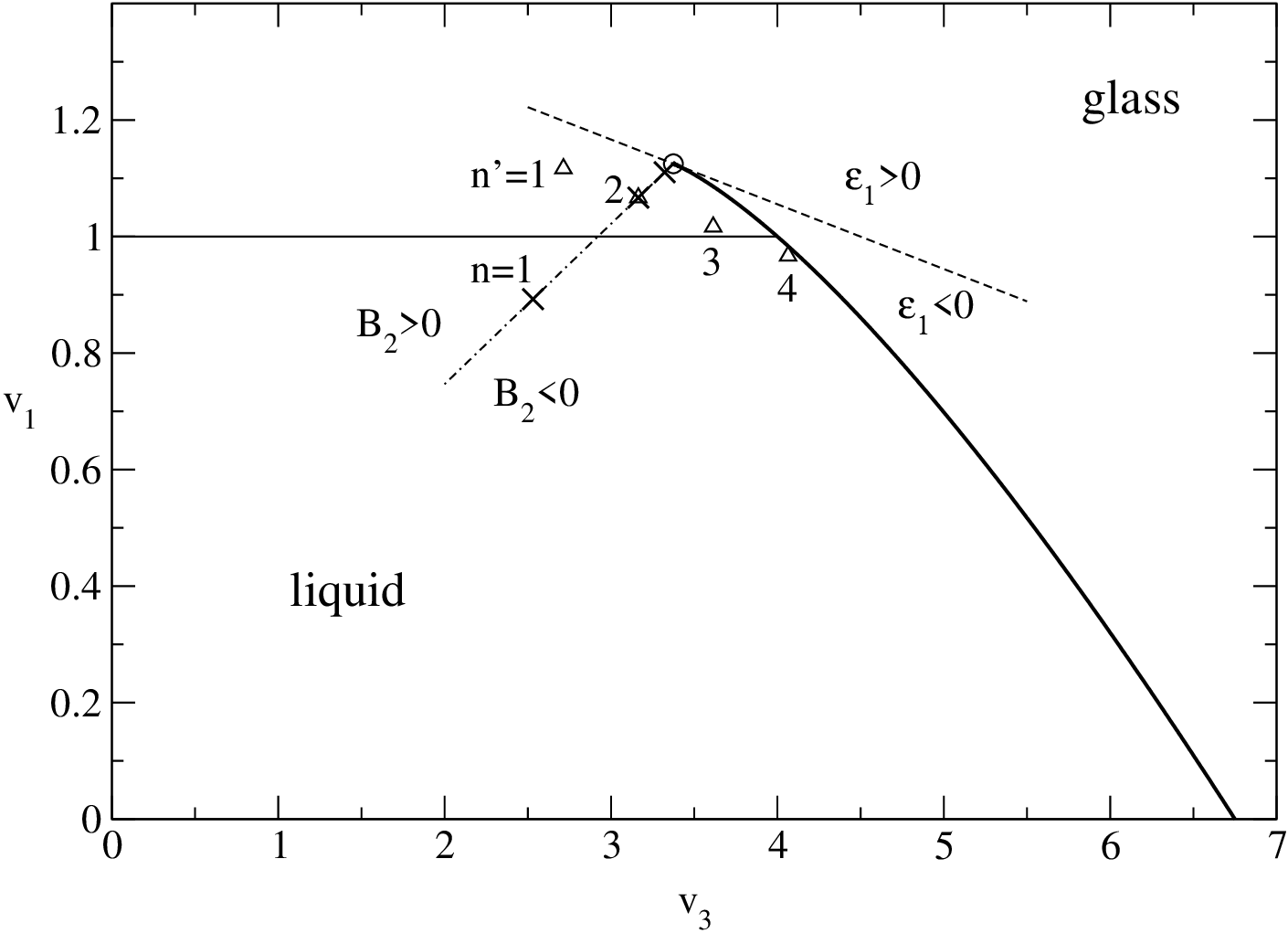}
\caption{\label{fig:phasediagram_F13}Phase diagram for the one-component
model defined in Sec.~\ref{sec:one_F13}. The horizontal light full line marks 
the liquid-glass-transition curve connected with a continuous variation of the 
glass-form factor. The heavy full line presents the set of $A_2$ singularities,
that ends at the $A_3$ singularity marked by a circle. The dashed straight 
line describes the points of vanishing separation parameter $\epsilon_1$ and
the dashed-dotted one the points of vanishing coefficients $B_2$. Crosses with 
labels $n$ and triangles with label $n'$ denote states discussed in 
Figs.~\ref{fig:A3path_F13}--\ref{fig:spectra_A3path_F13} and 
Figs.~\ref{fig:tangentpath_F13},~\ref{fig:tangentpath_F13_Fou}, respectively.
}
\end{figure}

\begin{figure}[Hhtb]
\noindent\includegraphics[width=0.42\textwidth]{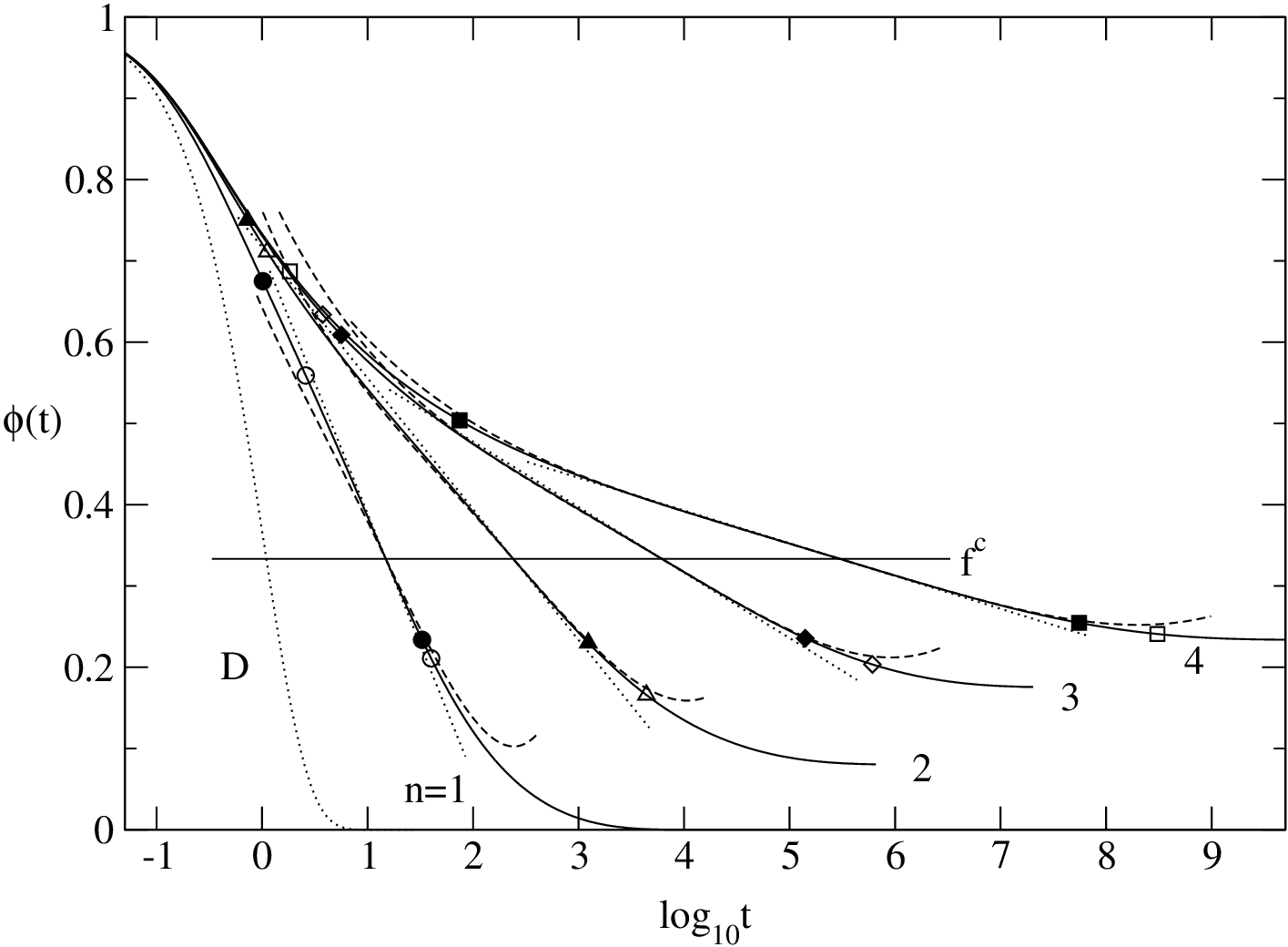}
\caption{\label{fig:A3path_F13}Correlators $\phi(t)$ for the one-component
model defined in Sec.~\ref{sec:one_F13}. The states are located on the line
$B_2=0$ with coupling constants: $v^c_1-v_1=0.9298/4^n$, $v^c_3-v_3=
3.3750/4^n$, $n=1,\dots, 4$, marked by crosses in 
Fig.~\ref{fig:phasediagram_F13}. The full lines are the solutions of 
Eqs.~(\ref{eq:F13_def}a, b) with $\tau_1=1$ as unit of time as also in the
following figures. 
The dotted straight lines exhibit the leading approximation, 
Eq.~(\ref{eq:phi_one_lead}), the dashed lines the leading correction, 
Eq.~(\ref{eq:phi_corr1}). The filled and open symbols, respectively, mark the
times where these approximations deviate from the solution by 5\%. 
The dotted line marked by 
D is the Debye law $\exp[-t/\tau_1]$.
The horizontal line shows the critical form factor $f^c=1/3$.
}
\end{figure}

The dominant deviation of the correlators from the logarithmic-decay law, 
Eq.~(\ref{eq:phi_one_lead}), is caused by the term $B_2\ln^2(t/\tau)$ in 
Eq.~(\ref{eq:phi_corr1}). Thus, the logarithmic-decay law is exhibited best 
for states $V$ with $B_2=0$. This line is shown dashed-dotted in 
Fig.~\ref{fig:phasediagram_F13}. Figure~\ref{fig:A3path_F13} demonstrates the
evolution of the dynamics upon shifting states on this line towards the $A_3$ 
singularity. The $\ln(t/\tau)$ interval, where Eq.~(\ref{eq:phi_one_lead}) or 
(\ref{eq:phi_corr1}) describe the correlators within an error margin of 5\% is
marked by closed or open symbols, respectively. For $n\geqslant 2$, these 
intervals increase with decreasing $V-V^c$.

There are two peculiarities concerning the range of applicability of the
asymptotic expansions. First, it can happen that for sufficiently large
$\epsilon$ the range shrinks if one proceeds from the leading approximation to 
the next-to-leading one as is demonstrated in Fig.~\ref{fig:A3path_F13} for the
$n=1,\,2$ results. This is caused by a cancellation of errors due to neglecting
the $B_1$-correction
in the prefactor of the  $\ln(t/\tau)$ term in Eq.~(\ref{eq:phi_corr1}) and due
to neglecting the terms proportional to $B_3$ and $B_4$. This peculiarity
would disappear if the tolerated error margin were decreased sufficiently
below the 5\% used. Second, for small $V-V^c$, the interval of decay for 
$\phi(t)$ below the critical form factor $f^c$ that is described by the 
asymptotic expansion shrinks with decreasing separation. This is inferred by
comparing the $n=3$ with the $n=4$ results. The reason is the following. The
correlator $\phi(t)$ decreases monotonically towards its long-time limit 
$f$ \cite{Goetze95b}. But the interval $f^c-f=-\hat{f}$ shrinks for 
$\epsilon\rightarrow 0$, since Eq.~(\ref{eq:A_l_mu_l}) implies
$-\hat{f} =(-\epsilon_1(V)/\mu_3)^{1/3}[1+{\cal O}(\epsilon^{1/3})]$.

Figure~\ref{fig:A3path_F13} demonstrates that the transient regime extends to
about $t/\tau_1=1$. For vanishing mode-coupling functional, the correlator
describes a Debye process: $\phi(t) = \exp(-t/\tau_1)$. Mode-coupling effects 
cause a slower decay for 
$t/\tau_1\geqslant 1$. But for $V$ close to $V^c$, the transient dynamics is 
rather insensitive to changes of the coupling constants. There is a crossover
interval, say $\tau_1<t<\tau^*$, before the decay of $\phi(t)$ towards $f^c$ 
can be described by the $\ln(t/\tau)$ law. The beginning $\tau^*$ of the range
of validity of Eq.~(\ref{eq:phi_one_lead}) is indicated by the filled symbols.
There are two subtleties demonstrated for $n\geqslant 2$. 
First, the time $\tau^*$ increases upon approaching the $A_3$
singularity, and therefore the decay interval $\phi(\tau^*)-f^c$ which is
described by the logarithmic law shrinks with decreasing separation parameters.
The control-parameter sensitive structural relaxation is governed by the two
time scales $\tau^*$  and $\tau$. Both times become large, but $\tau/\tau^*$ 
becomes large as well for $\epsilon\rightarrow 0$. Second, the beginning of the
range of applicability of the leading correction Eq.~(\ref{eq:phi_corr1}) is
control-parameter insensitive, as is shown by the open symbols on the 
short-time part of the decay curves.

\begin{figure}[Hhtb]
\noindent\includegraphics[width=0.42\textwidth]{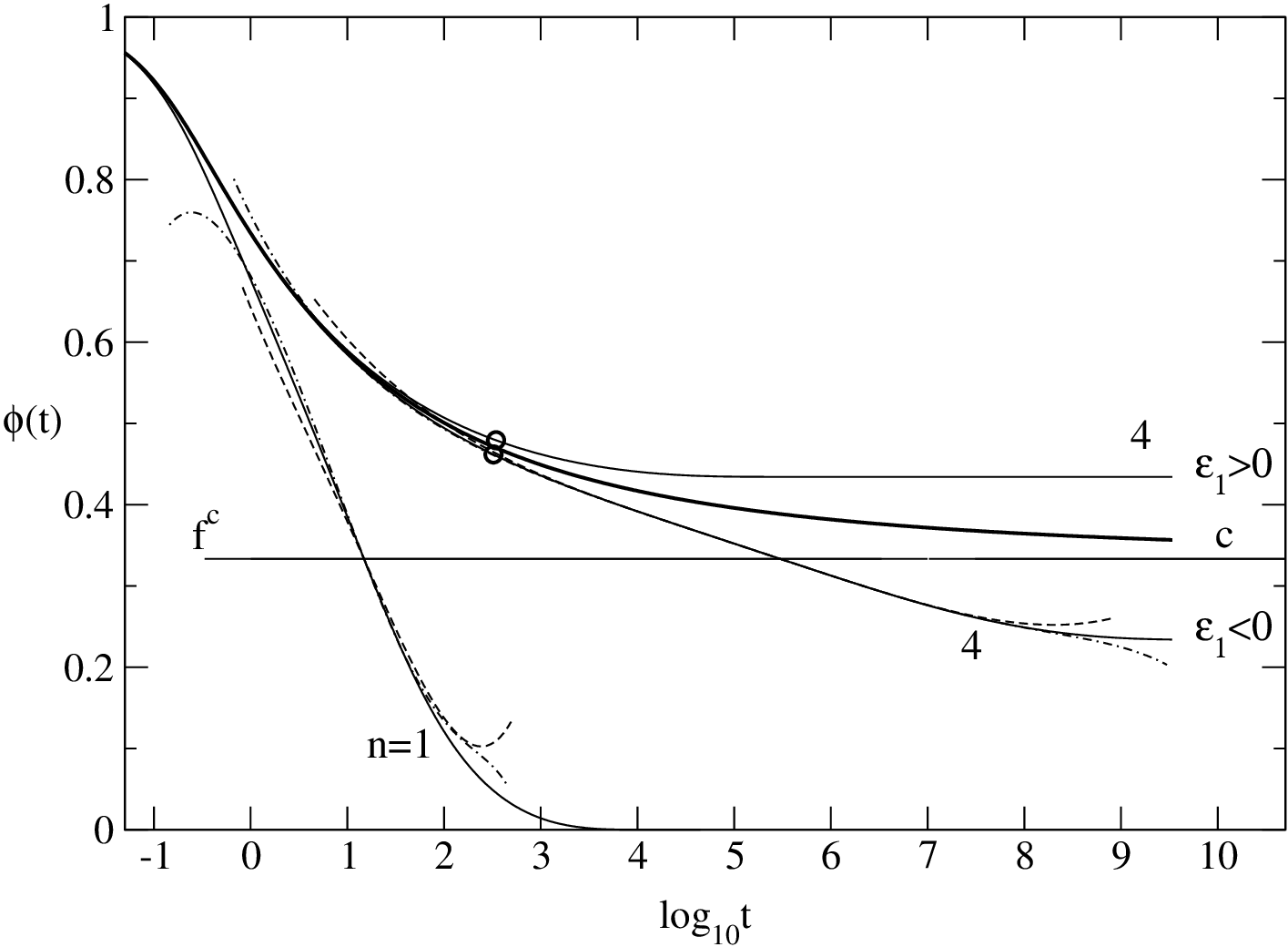}
\caption{\label{fig:critical_F13}The heavy full curve with label c is the 
solution of Eqs.~(\ref{eq:F13_def}a, b) at the $A_3$ singularity, $V=V^c$,
called the critical correlator.
The thin full curves with labels $n=1$ and
$4,\,\epsilon_1<0$ reproduce two of the solutions discussed in 
Fig.~\ref{fig:A3path_F13}. The dashed lines reproduce from 
Fig.~\ref{fig:A3path_F13} the corresponding approximations by the 
next-to-leading order, Eq.~(\ref{eq:phi_corr1}). The dashed-dotted curves 
extend
the asymptotic expansion by including the second order corrections,
Eq.~(\ref{eq:corr2}). The horizontal line marks the critical form factor 
$f^c=1/3$. The full curve marked with $n=4,\,\epsilon_1>0$ refers to a state,
$v_1-v_1^c=0.9298/4^4,\,v_3-v_3^c=3.3750/4^4$. The circles mark the times 
where the correlators $n=4,\,\epsilon_1\gtrless 0$ deviate by 2\% from the 
critical one.}
\end{figure}

The explanation of the findings in the preceding paragraph is based on the 
fact that for every fixed finite time interval, the MCT solutions are smooth 
functions of the control parameters \cite{Goetze95b}. Therefore, for 
$\epsilon$ tending to zero, the correlator for state $V$ has to approach the
correlator for state $V^c$, the so called critical correlator. The latter is
shown as the heavy full line with label $c$ in Fig.~\ref{fig:critical_F13}.
Thus, for every time interval $0\leqslant t\leqslant t_{\rm max}$ and every 
error margin, there exists an $\epsilon^*$, so that $\phi(t)$ agrees with the 
critical correlator within the error margin for all $|\epsilon|<\epsilon^*$
and all $0\leqslant t\leqslant t_{\rm max}$. This feature is demonstrated in 
Fig.~\ref{fig:critical_F13} by the two curves with label $n=4$ and 
$\epsilon_1\gtrless 0$. They refer to states with $v_1-v_1^c = 
\pm 0.9298/4^4$ and $v_3-v_3^c = \pm 3.3750/4^4$. These correlators are very 
close to the critical one for $t\leqslant t_{\rm max}$; $t_{\rm max}\approx 
325$ is indicated by open circles in Fig.~\ref{fig:critical_F13}.
The critical correlator does
not exhibit an $\ln(t/\tau)$-part. Thus, the time $\tau^*$ for the onset of the
description by Eq.~(\ref{eq:G1}) has to increase beyond any bound if $\epsilon$
tends to zero. The asymptotic expansion in Sec.~\ref{sec:one} was based on
$|\phi(t)-f^c|$ to be small. This condition is fulfilled for the critical
correlator if the time is sufficiently large, since $\phi(V^c,t)$
decreases monotonically to $f^c$. Hence, for 
$\epsilon\rightarrow 0$ there must appear the increasing time interval 
$\tau_1 < t < \tau^*$ where the asymptotic expansion describes the critical
correlator. The $\epsilon$-dependence due to the separation parameters
$\epsilon_n(V)$ and the $\epsilon$-dependence due to the time scale $\tau$
cancel to produce the critical correlator outside the transient regime and
prior to the onset of the $\ln(t/\tau)$ law. This is shown most clearly by the
dashed-dotted lines in Fig.~\ref{fig:critical_F13}. They exhibit the result of 
the asymptotic expansion up to the second correction given by 
Eq.~(\ref{eq:corr2}). They describe the complete decay for $t/\tau_1\geqslant 
1$ except for the final exponential approach towards 
$f=\phi(t\rightarrow\infty)$; and this for states as far from
the critical point as given by the one with label $n=1$ (cf. 
Fig.~\ref{fig:phasediagram_F13}).

\begin{figure}[Hhtb]
\noindent\includegraphics[width=0.42\textwidth]{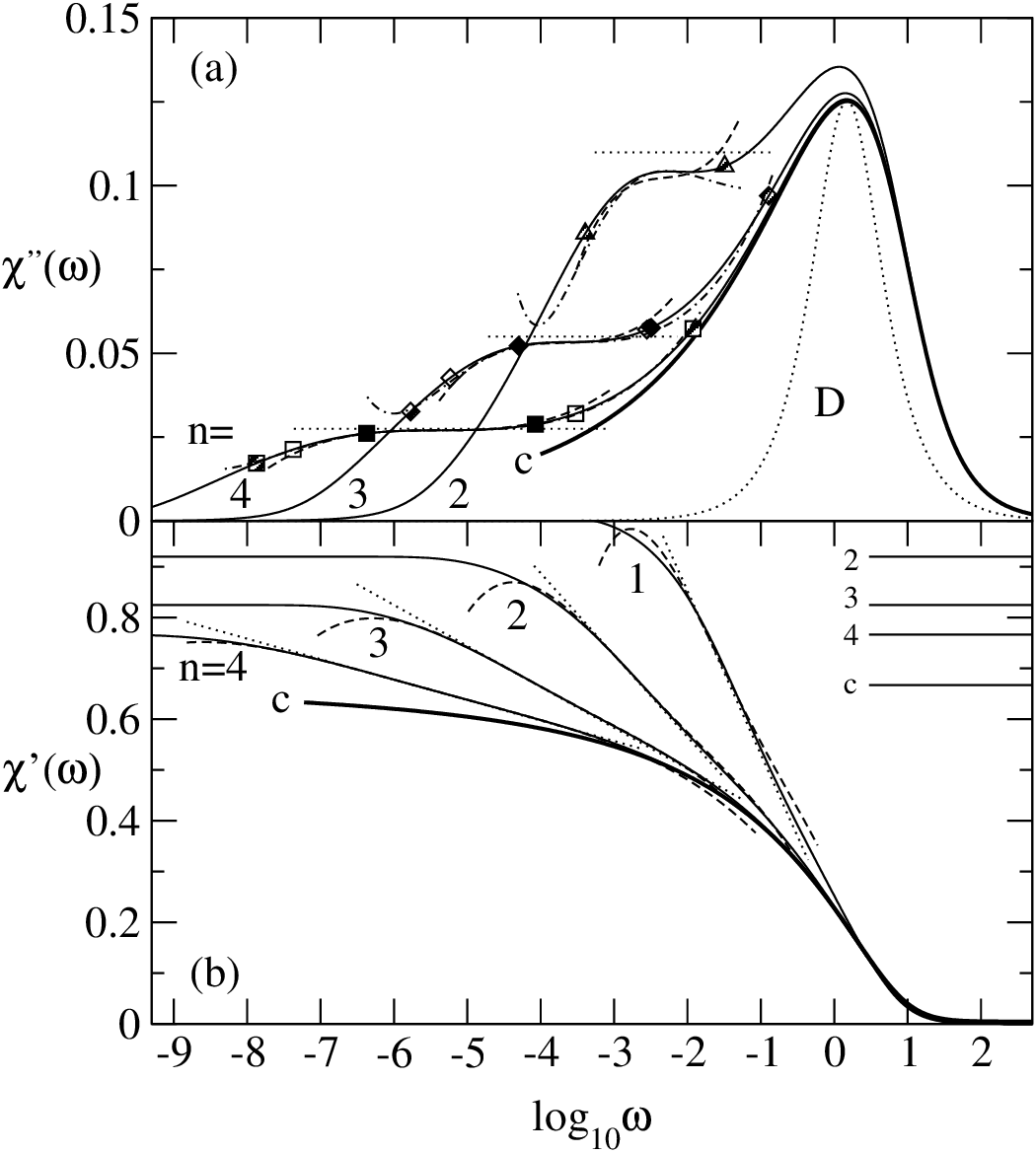}
\caption{\label{fig:spectra_A3path_F13}Susceptibility spectra $\chi''(\omega)$
and reactive parts of the dynamical susceptibility $\chi'(\omega)$ for the 
one-component model defined in Sec.~\ref{sec:one_F13}. The full lines with 
labels $n$ and $c$ correspond to the correlators with the same labels shown 
in Figs.~\ref{fig:A3path_F13} and \ref{fig:critical_F13}. The 
dotted straight lines, the dashed lines, and the dashed-dotted lines are the
leading approximation, Eq.~(\ref{eq:phi_one_lead}), the leading correction,
Eq.~(\ref{eq:phi_corr1}), and the second correction, Eq.~(\ref{eq:corr2}),
respectively. The filled, open, and half-filled symbols mark the 
frequencies where the corresponding approximation deviates from the spectrum
by 5\%. The dotted line with label D exhibits a Debye spectrum 
$C_D(\omega\tau_D)/[1+(\omega\tau_D)^2]$ with $C_D=0.2503$ and 
$\tau_D/\tau_1=0.670$
fitted to the maximum of the critical spectrum. The horizontal lines in the 
lower panel exhibit the static susceptibility $\chi'(\omega\rightarrow 0)=
1-f$ for the states with labels $n=2, 3, 4,$ and $c$, respectively.
}
\end{figure}

Figure~\ref{fig:spectra_A3path_F13} exhibits the dynamical susceptibilities
$\chi(\omega)=1-{\cal S}[\phi(t)](\omega+i 0)=\chi'(\omega)-i\chi''(\omega)$
for the states discussed above. Without interaction effects, the susceptibility
spectrum were a Debye peak $\chi''_D=C_D \omega \tau_D/[1+(\omega\tau_D)^2]$ 
with
$C_D=1,\,\tau_D=\tau_1$. Such Lorentzian spectrum is added to the upper panel
as dotted line with label $D$, where $C_D$ and $\tau_D$ are fitted to the 
maximum of the critical susceptibility spectrum. This shows, that the spectral
peaks near $\omega=1$, in particular their high-frequency wings, are due to the
transient dynamics. However, the low-frequency wings of the peaks are enhanced 
relative to the Debye spectrum and they are stretched to lower frequencies due 
to the critical relaxation 
within the interval $1/\tau^* < \omega < 1/\tau_1$. It is the structural
relaxation near the $A_3$ singularity which causes the skewed shape of the 
spectral peaks. The leading approximation, Eq.~(\ref{eq:phi_one_lead}), implies
constant-loss plateaus: $\chi''(\omega)/(1-f^c)=\pi B /2$.
 However, this formula describes the 
plateau only for $n\geqslant 3$ and then for a frequency interval that is
considerably smaller than the time interval for which 
Eq.~(\ref{eq:phi_one_lead}) describes the correlators in 
Fig.~\ref{fig:A3path_F13}. The leading corrections in 
Eq.~(\ref{eq:phi_corr1}) are much more important for an adequate description 
of the spectra than for the approximation of the correlators. The second
correction, Eq.~(\ref{eq:corr2}), is necessary to describe the plateau
for the $n=2$ state within a 5\% error margin. It is also necessary to describe
the crossover of the spectrum from the plateau towards the critical one for 
$n=3,\,4$.

To understand the range of validity of Eq.~(\ref{eq:phi_one_lead}) and its 
Fourier transform, one has to compare it with Eq.~(\ref{eq:phi_corr1}) and its 
Fourier transform, respectively. This amounts to comparing polynomials in
$\ln(t/\tau)$ and $\ln(i/\omega\tau)$. Let us restrict ourselves to the 
dominant terms for the model
to grasp the essence. Then one can write $(\phi(t)-f^c)/(1-f^c)=
-B\ln(t/\tau)[1-(B_4/B)\ln^3(t/\tau)]$. Thus, within the error margin $\delta$,
the leading linear-in-$\ln(t/\tau)$ approximation holds for $|\ln(t/\tau)|
\leqslant\sqrt[3]{\delta(B/B_4)}$. For the spectrum one gets from 
Eq.~(\ref{eq:Strafo_log}) in leading order $\chi''(\omega)/(1-f^c) = 
B(\pi/2)[1-4(B_4/B)\ln^3(1/\omega\tau)]$. Hence, the spectrum is at the plateau
within a deviation $\delta$ for $|\ln(1/\omega\tau)|\leqslant
\sqrt[3]{\delta(B/4B_4)}$. As a result, the range of applicability on a
logarithmic axis shrinks by a factor $\sqrt[3]{4}$ if one transforms from the 
time domain to the frequency domain.

Because of Eq.~(\ref{eq:Strafo_log}), 
${\cal S}[\ln^n(t/\tau)](z)=\ln^n(i/z\tau)-n\gamma\ln^{n-1}(i/z\tau)+\cdots$.
In leading approximation for $t\rightarrow\infty$ and $z\rightarrow 0$, 
one finds
$\chi'(\omega)\propto 1-\phi(1/\omega)$ whenever $\phi(t)$ is a polynomial in
$\ln(t/\tau)$. This explains the lower panel of 
Fig.~\ref{fig:spectra_A3path_F13} as a different representation of 
Figs.~\ref{fig:A3path_F13} and \ref{fig:critical_F13}. In particular, the 
linear-$\ln(\omega)$ parts in Fig.~\ref{fig:spectra_A3path_F13} are of a 
similar size as the linear-$\ln(t/\tau)$ parts in Fig.~\ref{fig:A3path_F13}.

\begin{figure}[Hhtb]
\noindent\includegraphics[width=0.42\textwidth]{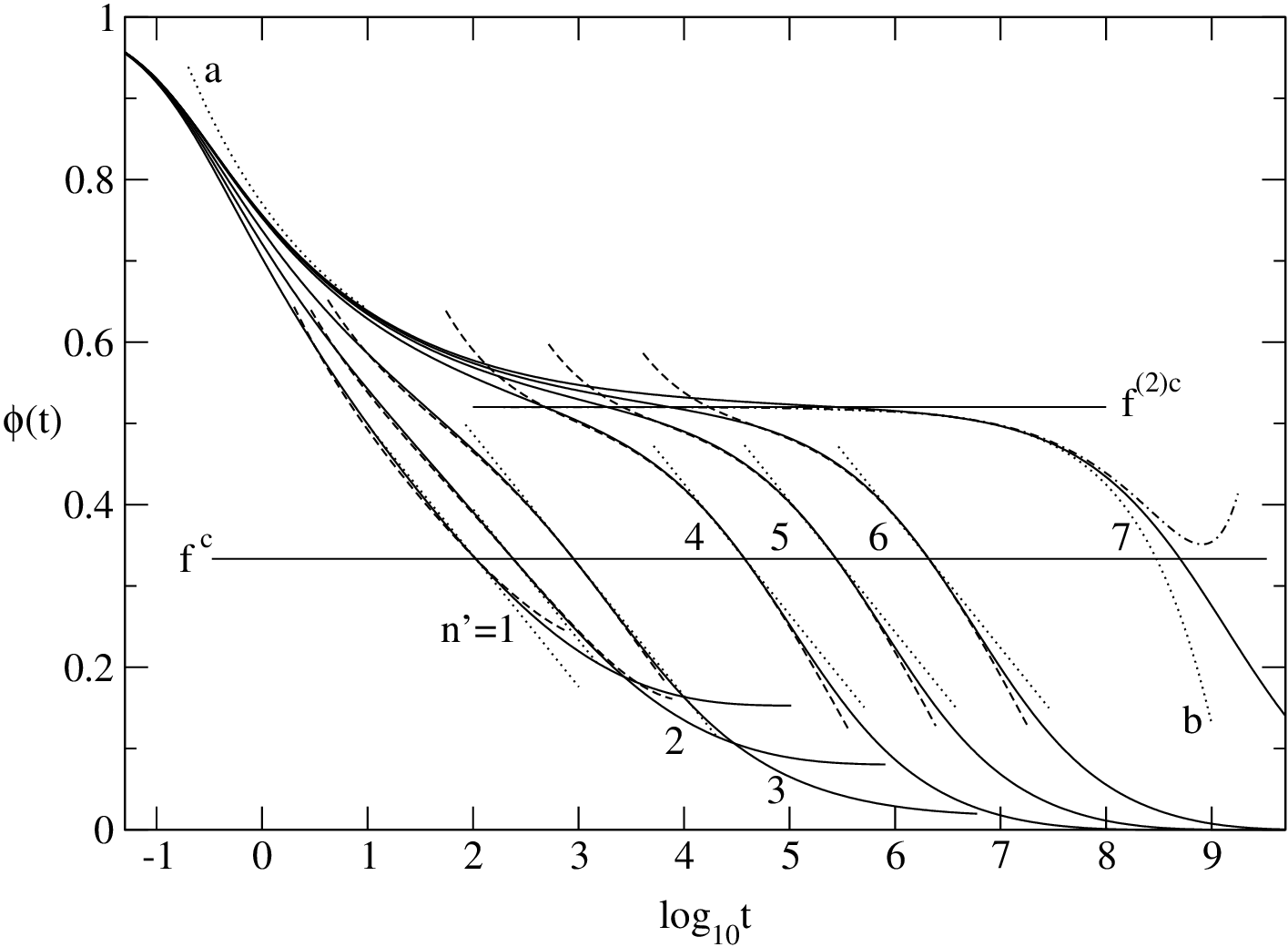}
\caption{\label{fig:tangentpath_F13}Correlators $\phi(t)$ for the one-component
model defined in Sec.~\ref{sec:one_F13} for states located on the line 
$\epsilon_1=-0.0182$. The states with labels $n'=1$--$4$ have the coordinates
$(v_1,v_3)=(1.1169,2.7141),\, (1.0669,3.1641),\, (1.0169,3.6141)$,
$(0.9669,4.0641)$, respectively, and they are marked by triangles in 
Fig.~\ref{fig:phasediagram_F13}. The state labeled $n'=2$ is identical with the
state discussed in Figs.~\ref{fig:phasediagram_F13}, \ref{fig:A3path_F13},
\ref{fig:spectra_A3path_F13} with label $n=2$. The states with labels $5$, $6$,
and $7$ have the coordinates $(0.9599,4.1271),\,(0.9569,4.1541)$, and 
$(0.9549,4.1721)$, respectively. The straight line through the states $1$--$7$
crosses the liquid-glass-transition curve at the state 
$V^{(2)c}=(0.95466,4.17407)$, where the critical glass-form factor has the 
value $f^{(2)c}=0.520$. The exponent parameter for the $A_2$-glass-transition
singularity is $\lambda=0.719$ implying a critical exponent $a=0.318$ and a 
von~Schweidler exponent $b=0.608$. The critical decay
law $\phi(t)-f^{(2)c}\propto t^{-a}$ and von~Schweidler's law  
$\phi(t)-f^{(2)c}\propto -t^b$ are shown by dotted lines labeled a
and b, respectively; the constants of proportionality are fitted to 
curve $7$. The dashed-dotted curve extends the von~Schweidler expansion for
curve $7$ to 
$\phi(t) = f^{(2)c} -(t/\tilde{\tau})^b+1.48(t/\tilde{\tau})^{2b}$.
The horizontal lines mark the critical glass-form factors $f^{(2)c}$
and $f^c$, respectively. The dotted straight lines and the dashed curves
are the leading asymptotic laws, Eq.~(\ref{eq:phi_one_lead}),  and the 
leading correction, Eq.~(\ref{eq:phi_corr1}), respectively.}
\end{figure}

Let us consider the states labeled $n'=1$--$3$ and shown by triangles in 
Fig.~\ref{fig:phasediagram_F13} in order to analyze the implications of the
correction term in Eq.~(\ref{eq:phi_corr1}) proportional to $B_2$. These states
are chosen on the line $\epsilon_1=-0.0182$ and state $n'=2$ is identical
with state $n=2$ considered in Fig.~\ref{fig:A3path_F13} as example for 
$B_2=0$. Figure~\ref{fig:tangentpath_F13} exhibits the correlators together 
with their approximations.
For $B_2>0$, the $\phi(t)$-versus-$\log(t)$ diagram is concave for all times 
outside the transient, since a parabola with positive curvature is added to the
leading linear variation described by Eq.~(\ref{eq:phi_one_lead}). The formula
with the leading correction describes the complete structural relaxation, 
except for the very last piece for the approach to the long-time limit $f$, as
shown by curve $n'=1$. This observation also holds for cases with $B_2<0$ as 
is demonstrated for state $n'=3$. However, for negative $B_2$, the 
$\phi(t)$-versus-$\log(t)$ curve exhibits two inflection points because 
$\phi(t)$ crosses the critical form factor $f^c$ with negative curvature. Since
the $\phi(t)$-versus-$\log(t)$ curve is convex for $\phi(t)\approx f^c$, it 
has to have an inflection point for 
$\phi(t)<f^c$ in order to approach the exponential, i.e. concave, long-time
asymptote. It has to exhibit an inflection point also for $\phi(t)>f^c$ in
order to approach the concave critical correlator for short times. The 
described alternation of convex and concave parts is identical to the 
behavior discussed earlier for the MCT correlators for states near an $A_2$
bifurcation \cite{Goetze91b,Franosch97}. 
But, contrary to the characteristic decay pattern
found for the MCT liquid-glass transition, curve $n'=3$ does not show
a two step relaxation scenario, even though there is a huge stretching
of the dynamics. For the decay from $0.80$ to $0.05$ a dynamical window of 
5~orders of magnitude is required. 
Within this large window, the correlator follows closely the law $\phi(t)
\propto \ln(t/\tau_{\rm eff})$.

The qualitative features described above for state $n'=3$ are more pronounced 
for state $n'=4$, since $B_2$ is decreased to larger negative values. The 
relaxation curve $4$ has the form expected for states near a liquid-glass 
transition. To corroborate this statement, further states $5$--$7$ are 
considered on the line $\epsilon_1=-0.0182$ between the state $4$ and the 
intersection $V^{(2)c}$ of this
line with the liquid-glass transition curve. The transition point $V^{(2)c}$ is
characterized by a critical glass-form factor $f^{(2)c}>f^c$. The decay of the
correlator from the value $f^{(2)c}$ to zero is the corresponding
$\alpha$-process. Its initial part is described by von Schweidler's power 
law, as indicated in Fig.~\ref{fig:tangentpath_F13} for curve $n'=7$ by the 
dotted line. In this case, von~Schweidler's law accounts for
the decay from $f^{(2)c}$ to about $0.45$, i.e. for about 15\% of the 
$\alpha$-relaxation. The analytical description of the $\alpha$-process can be
expanded by using the extension of von~Schweidler's law,
$\phi(t)-f^{(2)c}\propto -(t/\tilde{\tau})^b+\tilde{B}(t/\tilde{\tau})^{2b}$
\cite{Franosch97}, as shown by the dashed-dotted line.
Asymptotically, the $\alpha$-process obeys the 
superposition principle: $\phi(t)=\tilde{\phi}(t/\tau_\alpha)$, where 
$\tilde{\phi}$ is the control-parameter-independent shape function. 
The $\phi(t)$-versus-$\log(t)$ curves for the  $\alpha$-process can be 
superimposed by rescaling the time, i.e. by shifts parallel to the 
$\log(t)$-axis. The reader can check by herself that the curves $n'=4$--$7$ 
have 
the same shape for $\phi(t)<f^{(2)c}$. Outside the transient for 
$\phi(t)>f^{(2)c}$, the correlator follows the critical decay law for the fold 
bifurcation $\phi(t)-f^{(2)c}\propto 1/t^a$, as is also demonstrated for curve 
$7$. The results for states 
$4$--$7$ exemplify the well understood scenario for the evolution of structural
relaxation near a liquid-glass transition. The formula~(\ref{eq:phi_corr1}) 
provides an accurate description of 60\% of the $\alpha$-process.

Comparison of the results for states $n'=1$--$3$ with the second-correction 
formula based on Eq.~(\ref{eq:corr2}) yields the same conclusions as discussed
above in connection with Fig.~\ref{fig:critical_F13}. The second-correction 
formula does 
not alter seriously the fit quality for the long-time part of the curves 
$n'=4$--$6$ in  Fig.~\ref{fig:tangentpath_F13}. However, for $\phi(t)\approx 
f^{(2)c}$, the extended formula yields slightly worse results than 
Eq.~(\ref{eq:phi_corr1}). This is so, since for $\phi(t)\gtrsim f^{(2)c}$, the
dynamics is governed by the $A_2$ singularity $V^{(2)c}$ whose existence is 
ignored in the expansions near the higher-order singularity $V^c$. The number
$f^{(2)c}-f^c$ marks the limit where the expansion in the small parameter
$\phi(t)-f^c$ makes sense. The opposite conclusion holds for the description
of the $\alpha$-process for $\phi(t)\approx f^c$. Von~Schweidler's law results
from an expansion for states $V$ near $V^{(2)c}$
in terms of the small parameter $f^{(2)c}-\phi(t)$. This number becomes too 
large if $\phi(t)\approx f^c$. It is the dynamics dominated by the higher order
glass-transition singularity $V^c$ that ruins the relevance of the expansion 
resulting in the von~Schweidler law. The stretching of the $\alpha$-process
connected with the transition of $V^{(2)c}$ is larger than estimated by 
von~Schweidler's law, because of the logarithmic-decay effects.

\begin{figure}[Hhtb]
\noindent\includegraphics[width=0.42\textwidth]{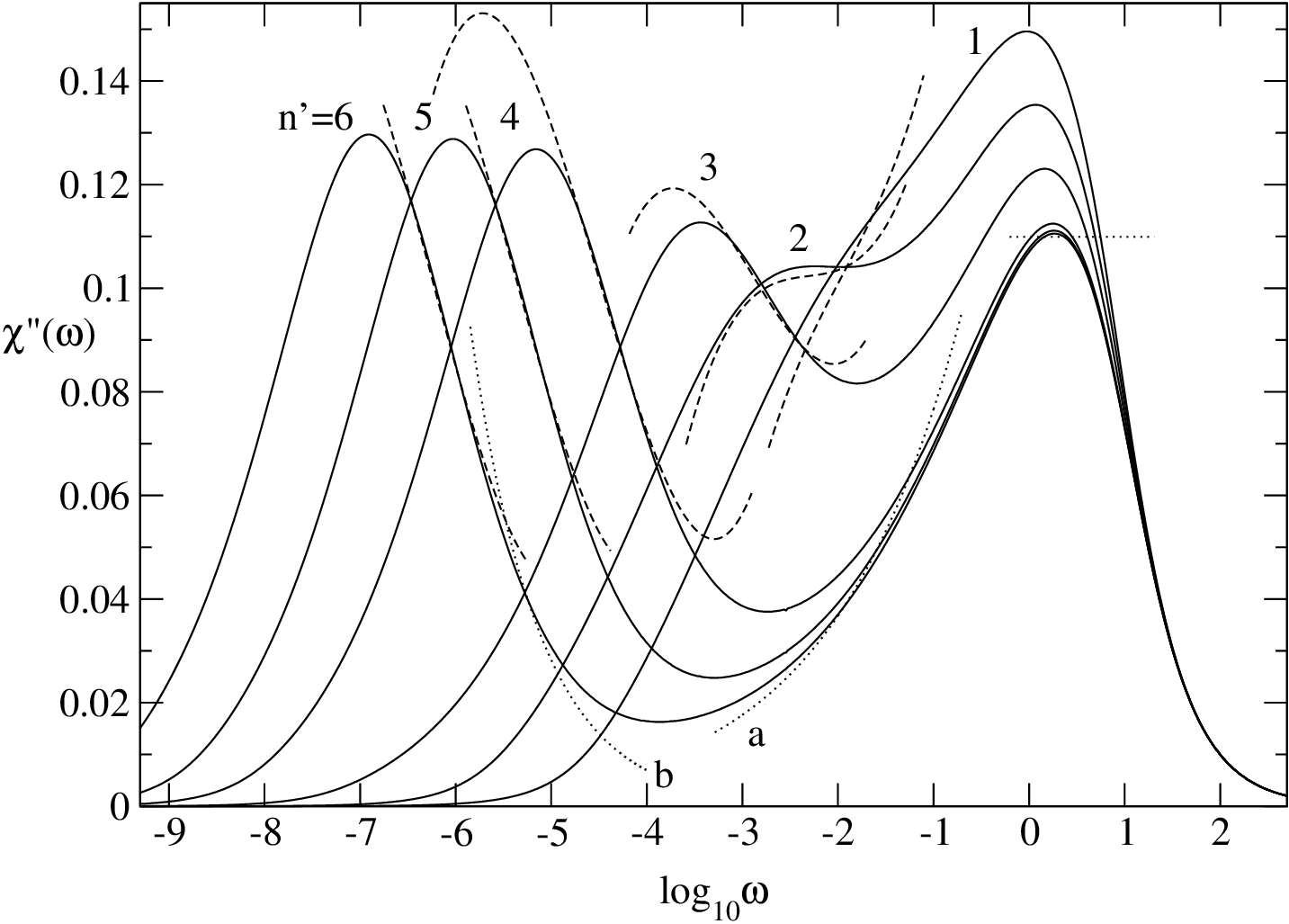}
\caption{\label{fig:tangentpath_F13_Fou}Susceptibility spectra $\chi''(\omega)$
for the correlators in Fig.~\ref{fig:tangentpath_F13} for the labels 
$n'=1$--$6$. The dashed lines are obtained from the leading-correction 
formula, 
Eq.~(\ref{eq:phi_corr1}). The dotted line with label a is the critical spectrum
proportional to $\omega^a$ and dotted line with label b is the von~Schweidler
law $\chi''(\omega)\propto\omega^{-b}$ for state $n'=6$. The dotted horizontal
line corresponds to the spectrum of the leading approximation, 
Eqs.~(\ref{eq:G1}, \ref{eq:phi_one_lead}), $\chi''(\omega)=(1-f^c)
\sqrt{-6\epsilon_1}$, that is shared by all states $n'=1$--$6$.}
\end{figure}

Figure~\ref{fig:tangentpath_F13_Fou} exhibits the susceptibility spectra
calculated from correlators discussed in Fig.~\ref{fig:tangentpath_F13}. The 
results for the states $n'=1,\,2,\,$ and $3$ exhibit the evolution of an
$\alpha$-peak if the states cross the line $B_2=0$ in the phase diagram of 
Fig.~\ref{fig:phasediagram_F13}. The leading-correction formula, 
Eq.~(\ref{eq:phi_corr1}), describes this scenario qualitatively. The spectra 
for states $n'=4,\,5,\,$ and $6$ exhibit the superposition principle for the 
$\alpha$-peak of the susceptibility spectra. The $A_3$-dynamics causes a
high-frequency wing of the $\alpha$-peak following closely a linear variation
with the logarithm of the frequency: $\chi''(\omega)\propto -\ln (\omega\tau)$.
This phenomenon is described well by Eq.~(\ref{eq:phi_corr1}) and it causes a
strong $\alpha$-process stretching. The $\alpha$-peak width at half of the
$\alpha$-peak hight is about $2.5$ decades. Von~Schweidler's asymptotic law is
irrelevant for the description of the $\alpha$-peak for states $n'=4$--$6$.

\section{\label{sec:q}Relaxation Formulas for States Near Higher-Order 
Glass-Transition Singularities}

In this section, the generalizations of Eqs.~(\ref{eq:phi_one_lead}) and 
(\ref{eq:phi_corr1}) shall be derived for the asymptotic expansion of the 
solutions dealing with an arbitrary number $M$ of correlators. In particular, 
the general formulas for $\mu_2$, $\mu_3$, $\zeta$ and for the 
separation parameters $\epsilon_1(V)$ and $\epsilon_2(V)$ shall be obtained. 
The starting formulas are Eqs.~(\ref{eq:J_def}) and (\ref{eq:J_exp}).  
The subtlety of the problem is the treatment of the singular 
$M$ by $M$ matrix $[\delta_{qk} - A_{qk}^{(1)c}]$.

\subsection{\label{subsec:q_asy}Asymptotic expansion of the 
equations of motion}

The left and right eigenvectors of matrix $A_{q k}^{(1) c}$ for
the maximum eigenvalue $E^c = 1$ shall be denoted by $a^*_k$
and $a_k, \, k = 1, \ldots , M$, respectively. According to the
Frobenius theorems \cite{Gantmacher74b}, one can require $a^*_k
\geqslant 0$ and $a_k \geqslant 0$. It will be convenient to fix the
eigenvectors uniquely by the conditions
$\sum_q \, a^*_q \, a_q = 1$ and $\sum_q \, a^*_q \, a_q^2 =
1$. The solubility condition of Eq.~(\ref{eq:J_def}) reads
\begin{subequations}\label{eq:qsol_general}
\begin{equation}\label{eq:J_solubility}
\sum_q a^*_q J_q (z) = 0 \,\, ,
\end{equation}
and its general solution can be written as
\begin{equation}\label{eq:phisol}
\hat \phi_q (t) =   a_q \hat \phi (t) + \tilde \phi_q (t) \,\, .
\end{equation}
The splitting of $\hat \phi_q (t)$ in two terms is unique if one imposes the 
condition $\sum_q \, a^*_q \, \hat \phi_q (t) = \hat \phi (t)$. The part 
$\tilde\phi_q (t)$ can be expressed by means of the reduced resolvent $R_{q k}$
of $A_{q k}^{(1) c}$:
\begin{equation}\label{eq:a_perpendiclar}
{\cal S}[ \tilde \phi_q (t)] (z) = R_{q k} J_k (z)
\,\, .
\end{equation}\end{subequations}
It is an elementary task to evaluate from matrix $A_{q k}^{(1) c}$ the vectors 
$a^*_k,\,a_k$ and the matrix $R_{qk}$ \cite{Gantmacher74b}.

Equations~(\ref{eq:J_exp}) and~(\ref{eq:qsol_general}) suggest an expansion of 
$\hat{\phi}(t)$ as Eq.~(\ref{eq:solution_expansion}) and
\begin{subequations}\label{eq:phitilde}
\begin{equation}\label{eq:phiq_Gi}
  \tilde{\phi}_q (t)    = G^{(2)}_q (t) + G^{(3)}_q (t) + \cdots \,\,
  , \quad G^{(n)}_q (t) = {\cal O} (|\epsilon|^{n/2})
\,\, ,
\end{equation}
\begin{equation}\label{eq:Ji_expansion}
J_q (t) = J^{(2)}_q(t) + J^{(3)}_q (t) + \cdots \,\,, \quad
	J^{(n)}_q (t) = {\cal O} (|\epsilon|^{n/2})\,\,.
\end{equation} 
\end{subequations}
Here, for example, 
\begin{subequations}\label{eq:Ji}
\begin{eqnarray}\label{eq:Ji_J2}
J^{(2)}_q (z) & = & \hat{A}_q^{(0)} (V)  +  A_{q
k_1 k_2}^{(2)c} a_{k_1}  a_{k_2} {\cal S}[{G^{(1)}}^2 (t)](z) 
\nonumber \\&&
- a_q^2 {\cal S}[G^{(1)} (t)]^2 (z) \,\, ,
\end{eqnarray}
\begin{eqnarray}\label{eq:Ji_J3}
J^{(3)}_q (z) & = & 2 \{  A_{q k_1 k_2}^{(2)c} a_{k_1}  a_{k_2}
{\cal S} [G^{(1)} (t) G^{(2)} (t) ] (z) 
	\nonumber \\ &&\qquad - a_q^2 {\cal S}
[G^{(1)} (t) ] (z) {\cal S} [  G^{(2)} (t) ] (z) \}
\nonumber \\ 
& + & \hat{A}_{q k}^{(1)} (V) a_k {\cal S} [G^{(1)}
(t) ] (z) \nonumber \\ & + & 2 \{  A_{q k_1 k_2}^{(2)c}
a_{k_1} {\cal S} [G^{(1)} (t) G^{(2)}_{k_2} (t) ] (z) 
	\nonumber \\ &&\qquad - a_q
{\cal S} [G^{(1)} (t) ] (z) {\cal S} [  G^{(2)}_q (t)
] (z) \} \nonumber \\ 
& + & A_{q k_1 k_2 k_3}^{(3)c} a_{k_1}
a_{k_2} a_{k_3} {\cal S} [{G^{(1)}}^3 (t) ] (z) 
	\nonumber \\ &&\qquad - a_q^3 {\cal
S} [G^{(1)} (t) ]^3 (z) \,\, .
\end{eqnarray}
\end{subequations}
The justification of the preceding expansions shall be given by demonstrating 
how the equations can be solved recursively.

\subsection{\label{subsec:q_lead}The leading-order contribution}

The leading-order contribution to the solubility condition is obtained by 
substituting Eq.~(\ref{eq:Ji_J2}) into Eq.~(\ref{eq:J_solubility}). One 
arrives at: $\epsilon_1(V)+\lambda {\cal S} [{G^{(1)}}^2(t)] (z)
- {\cal S} [G^{(1)} (t)]^2 (z) = 0$. Here $\lambda = \sum_q \, a^*_q \, 
A_{q k_1 k_2}^{(2) c} \, a_{k_1} \, a_{k_2}$ is the expression for the 
exponent parameter \cite{Goetze91b,Franosch97} and
\begin{equation}\label{eq:epsilon1_A}
 \epsilon_1 (V) = \sum_q a^*_q \hat{A}_q^{(0)} (V)\,\,.
\end{equation}
The $z=0$ limit leads to $\epsilon_1(V)+(\lambda-1){\hat{f}}^{(1)}{}^2=0$.
Comparison with Eq.~(\ref{eq:A_l_mu_l}) yields the conclusion that 
\begin{equation}\label{eq:mu2_from_lambda}
\mu_2 = 1 - \sum_q \, a^*_q \, A_{q k_1 k_2}^{(2) c} \, a_{k_1} \, a_{k_2}
\,\, .
\end{equation}
This parameter has to be zero according to Eq.~(\ref{eq:lambda_unity})
in order for $V^c$ to be a higher-oder singularity. For $\mu_2=0$, $\lambda=1$,
and the equation found for $G^{(1)}(t)$ is identical with 
Eq.~(\ref{eq:solution_lead}). Thus, $\epsilon_1(V)$ is the first separation 
parameter and Eqs.~(\ref{eq:G1}) and (\ref{eq:eps1_negativ}) remain valid. 

Introducing the critical amplitude $h_q$ by the same formula as in the 
theory for the $A_2$-singularity \cite{Goetze91b,Franosch97}
\begin{equation}\label{eq:amplitude_h}
h_q = ( 1 - f_q^c) a_q \,\, ,
\end{equation}
the leading approximation for the correlators is 
\begin{equation}\label{eq:log_decay}
\phi_q (t) = f_q^c + h_q \left[ - B \ln (t / \tau) \right]  \,\, .
\end{equation}
Here, $B=\sqrt{-6\epsilon_1(V)}/\pi={\cal O}(|\epsilon|^{1/2})$.
Equation~(\ref{eq:log_decay}) describes the dynamics up to errors of order 
$\epsilon$; it is the generalization of the logarithmic decay law 
\cite{Goetze88b} to arbitrary MCT models.

Equation~(\ref{eq:log_decay}) is the factorization theorem of MCT. 
In leading order, $\phi_q(t)-f_q^c$ factorizes in two terms. The factor $h_q$
is time- and control-parameter-independent and it characterizes via its 
$q$-dependence the specific correlator. The other factor is the function
$G^{(1)}(t)=-B\ln(t/\tau)$. This factor is shared by all correlators. It 
describes via $B$ and $\tau$ the control-parameter dependence and via $\ln(t)$
the complete time dependence. Within the range of validity of 
Eq.~(\ref{eq:log_decay}), the rescaled correlators $\hat{\phi}_q(1) = 
[\phi_q(t)-f_q^c]/h_q$ are the same for all $q$.  Let us emphasize that
Eq.~(\ref{eq:log_decay}) is an exact limit result for the solutions of
Eqs.(\ref{eq:mct}) and (\ref{eq:mct_kernel}):
\begin{equation}\label{eq:asy_result}
\lim_{\epsilon\rightarrow 0}[\phi_q(\tilde{t}\tau)-f_q^c]
	/\sqrt{-\epsilon_1(V)} = -\sqrt{6/\pi^2} h_q \ln (\tilde{t})\,\,.
\end{equation}
The interval of rescaled times $\tilde{t}=t/\tau$, where 
$[\phi_q(t)-f_q^c]/\sqrt{-\epsilon_1(V)}$ becomes close to the r.h.s of 
Eq.~(\ref{eq:asy_result}), expands beyond any bound if $V$ approaches $V^c$ 
arbitrarily close. It will be shown below, how the leading corrections
for $\phi_q(t)$ describe violations of the factorization theorem.

Substitution of Eq.~(\ref{eq:Ji_J2})
into Eq.~(\ref{eq:a_perpendiclar}) yields the leading order contribution to 
$\tilde \phi_q (t)$, i.e. the function $G^{(2)}_q(t)$ in 
Eq.~(\ref{eq:phiq_Gi}). Equation~(\ref{eq:solution_lead}) is used to express 
${\cal S} [G^{(1)}(t)]^2 (z)$ in terms of ${\cal S} [{G^{(1)}}^2 (t)] (z)$ so 
that
\begin{subequations}\label{eq:G2q}
\begin{equation}\label{eq:G2q_Y}
 G^{(2)}_q (t) = X_q {G^{(1)}}^2 (t) + \hat{Y}_q (V) \,\, .
\end{equation}
The amplitude $X_q$ is independent of $\epsilon$,
\begin{equation}\label{eq:G2q_Xq}
X_q = R_{q k} \left[ A_{k k_1 k_2 }^{(2)c} a_{k_1} a_{k_2} -
a_{k}^2 \right] \,\, .
\end{equation}
$\hat{Y}_q (V) = {\cal O} (\epsilon)$ and reads
\begin{equation}\label{eq:G2q_Rqk}
\hat{Y}_q  (V)= R_{q k} \left[ \hat{A}_{k }^{(0)} (V) -
\epsilon_1 (V) a_k^2 \right] \,\, .
\end{equation}
\end{subequations}

\subsection{\label{subsec:q_corr1}The leading correction}

If one substitutes Eq.~(\ref{eq:G2q_Y}) into Eq.~(\ref{eq:Ji_J3}), one gets an
expression for $J^{(3)}_q (z)$ in terms of the known $G^{(1)}(t)$ and 
the unknown $G^{(2)} (t)$. Therefore, the solubility condition, 
Eq.~(\ref{eq:J_solubility}), evaluated up to order $\epsilon^{3/2}$, 
yields an equation for $G^{(2)} (t)$. The latter has the form of 
Eq.~(\ref{eq:problem_corr}), where also the inhomogeneity is given by 
Eq.~(\ref{eq:f2}). This holds with the formula
\begin{eqnarray}\label{eq:epsilon2_A}
\epsilon_2 (V) & = & \sum_q a^*_q \hat{A}_{q k}^{(1)} (V) a_k
+ 2 \epsilon_1 (V) \sum_q a^*_q a_q X_q \nonumber
\\ & + &  2 \sum_q a^*_q \left[ A_{q k_1 k_2}^{(2)c} a_{k_1} 
\hat{Y}_{k_2} (V) - a_q \hat{Y}_q (V) \right]  \,\,
\end{eqnarray}
for the second separation parameter, and the constants
\begin{equation}\label{eq:zeta}
\zeta = \sum_q a^*_q \left[ a_q X_q +  a_q^3/2
\right] \,\, ,
\end{equation}
\begin{equation}\label{eq:mu3}
\mu_3 = 2 \zeta - \sum_q a^*_q \left[ A_{q k_1 k_2 k_3}^{(3)c}
a_{k_1} a_{k_2} a_{k_3} + 2  A_{q k_1 k_2}^{(2)c} a_{k_1}
X_{k_2}
 \right] \,\, .
\end{equation}
As a result, Eqs.~(\ref{eq:Bcoeffs}) for the function $G^{(2)} (t)$
remain valid.

Combining the results for $G^{(1)}(t),\,G^{(2)}(t)$ and $G^{(2)}_q(t)$
with Eq.~(\ref{eq:phiq_Gi}), and this with Eq.~(\ref{eq:hat_phi}), one obtains 
the main result of this paper. It describes the correlators up to errors of 
order $|\epsilon|^{3/2}$:
\begin{eqnarray}\label{eq:G1G2q}
\phi_q (t) = (f_q^c + \hat{f}_q) &+& h_q  \bigl [  (- B + B_1)
\ln (t / \tau)  \nonumber
\\ && +  (B_2 + K_q B^2) \ln^2 (t / \tau)\nonumber
\\ && +     B_3 \ln^3 (t / \tau) +  B_4 \ln^4 (t / \tau)  \bigr ] \,\, .
\end{eqnarray}
Here
\begin{equation}\label{eq:deltafq}
\hat{f}_q = (1 - f_q^c) \hat{Y}_q
\end{equation}
is a renormalization of the glass-form factor of order $\epsilon$ to be 
calculated from Eq.~(\ref{eq:G2q_Rqk}). The critical 
amplitude $h_q$ is defined by Eq.~(\ref{eq:amplitude_h}). The parameter $B_1$ 
from Eq.~(\ref{eq:Bcoeffs_B1}) is a renormalization of order $\epsilon$ of the 
prefactor of the logarithmic decay law. The three terms  proportional to 
$B_2,\,B_3$, and $B_4$, respectively, describe leading deviations from the
logarithmic decay. They are of order $\epsilon$ and follow from 
Eqs.~(\ref{eq:Bcoeffs_B2}) and (\ref{eq:Bcoeffs_B3}). The relative size of 
these deviations is the same for all correlators. This means, that these terms
imply a modification of the factorization theorem, $\phi_q(t)-(f_q^c+\hat{f}_q)
= h_q G(t)$, in the sense that $G(t)=-B\ln(t/\tau)$ in Eq.~(\ref{eq:log_decay})
is to be generalized by the bracket on the r.h.s of Eq.~(\ref{eq:phi_corr1}).
It is solely the  contribution proportional to $B^2=-6\epsilon_1/\pi^2=
{\cal O}(\epsilon)$ which describes a violation of the factorization theorem. 
It enters with the correction amplitude
\begin{equation}\label{eq:Kq}
K_q = X_q / a_q  \,\, .
\end{equation}
Its $q$-dependence expresses the fact that the size of the leading corrections
depends on the chosen correlator. Thus, the range of validity of the universal
Eq.~(\ref{eq:log_decay}) is not universal. The correction amplitude is to be 
calculated from Eq.~(\ref{eq:G2q_Xq}). The formula for $K_q$ is the same as 
discussed in the theory for the $A_2$ singularity \cite{Franosch97}.

\section{\label{sec:BK}Results for a two-component model}

The simplest example exhibiting a generic swallow-tail bifurcation is given by 
an $M=2$ model with the mode-coupling functionals ${\cal F}_1[V,\tilde{f}_k] = 
v_1\tilde{f}_1^2 + v_2\tilde{f}_2^2,\,{\cal F}_2[V,\tilde{f}_k] = v_3
\tilde{f}_1\tilde{f}_2$.  This model was motivated 
originally as truncation of microscopic equations of motion for a 
symmetric molten salt \cite{Bosse87b}. The model shall be used here in order
to demonstrate implications of our theory which could not be demonstrated for 
the $M=1$ model studied in Sec.~\ref{sec:one_F13}. Using Brownian 
microscopic dynamics, the equations of motion~(\ref{eq:mct_eomBrown}) and 
(\ref{eq:mct_kernel}) read for $\phi_q(t),\,q=1,\,2$:
\begin{subequations}
\label{eq:BK_eom}
\begin{eqnarray}\label{eq:BK_eom_int}
&&\tau_q \partial_t \phi_q(t) + \phi_q (t) +
\int_0^t m_q (t - t^\prime) \partial_{t^\prime} \phi_q
(t^\prime)dt^\prime = 0\,,\\
\label{eq:BK_eom_m1}
&&m_1(t) = v_1 \phi_1^2(t) + v_2 \phi_2^2(t)\,,\\
\label{eq:BK_eom_m2}
&&m_2(t) = v_3 \phi_1(t)\phi_2(t)\,\,.
\end{eqnarray}
\end{subequations}
The three coupling constants $v_n\geqslant 0$ shall be considered as
the components of the control-parameter vector $V=(v_1, v_2, v_3)$.

Let us note convenient equations for the discussion of the phase diagram
\cite{Goetze88b,Goetze91b}, restricting ourselves  to $v_3>4$. 
Equation~(\ref{eq:singularity}) for the second form factor implies 
$f_2=[v_3f_1 - 1]/(v_3 f_1)$, and this result can be used to eliminate $f_2$
in the following expressions. Thus, Eq.~(\ref{eq:singularity}) for the first
form factor, $f_1/(1-f_1) = v_1 f_1^2 + v_2 f_2^2$, is a linear equation for 
$(v_1, v_2)$ with coefficients that are nonlinear in $f_1$ and $v_3$. The same
statement holds for Eq.~(\ref{eq:eigenvalue})
for a singularity which is equivalent to 
$f_1^{(2)c}/(1-f_1^{(2)c})^2 = 2 v_1^{(2)c} {f_1^{(2)c}}^2 + 2 {v_2^{(2)c}}^2 
f_2^{(2)c}(1-f_2^{(2)c})$. These equations can be used to express $v_1^{(2)c}$
and $v_2^{(2)c}$ in terms of $v_3^{(2)c}$ and $f_1^{(2)c}$. To ease the 
notation, variables $x$ and $y$ shall be introduced as
\begin{subequations}\label{eq:BK_PD}
\begin{equation}\label{eq:BK_PD_xy}
v_3^{(2)c} = x\,,\quad f_1^{(2)c} = y\,\,.
\end{equation}
One gets
\begin{eqnarray}\label{eq:BK_PD_v1}
v_1^{(2)c} &=& \frac{3 - (2+x) y}{2 (1-y)^2 y (2-x y)}
\,,\\\label{eq:BK_PD_v2}
v_2^{(2)c} &=&  \frac{x^2 y (y^2 - 2 y^3)}{2 (1-y)^2 (x^2 y^2 -3 x y + 2)}
\,\,.
\end{eqnarray}
These equations define the surface of bifurcation singularities of 
Eq.~(\ref{eq:singularity}) in the 3-dimensional parameter space. The variables
$x$ and $y$ with
$4<x$ and $1/2 \leqslant y \leqslant 3/(2+x)$ serve as surface parameters. The
exponent parameter $\lambda=1-\mu_2$ is determined by
\begin{equation}\label{eq:BK_PD_mu2}
\mu_2 = 
\frac{(3 x^2 + 6 x)y^3 - (x^2+18x+8)y^2 + (6x+18)y-6}
                {(2x^2+4x)y^3 - 12 x y^2 + (2x+4)y }
\,\,.
\end{equation}
\end{subequations}
 The maximum 
theorem, mentioned above in connection with Eq.~(\ref{eq:singularity}), has to
be used to identify among the points $(v_1^{(2)c}, v_2^{(2)c}, v_3^{(2)c})$ 
those that are glass-transition singularities. 

\begin{figure}[Hhtb]
\noindent\includegraphics[width=0.42\textwidth]{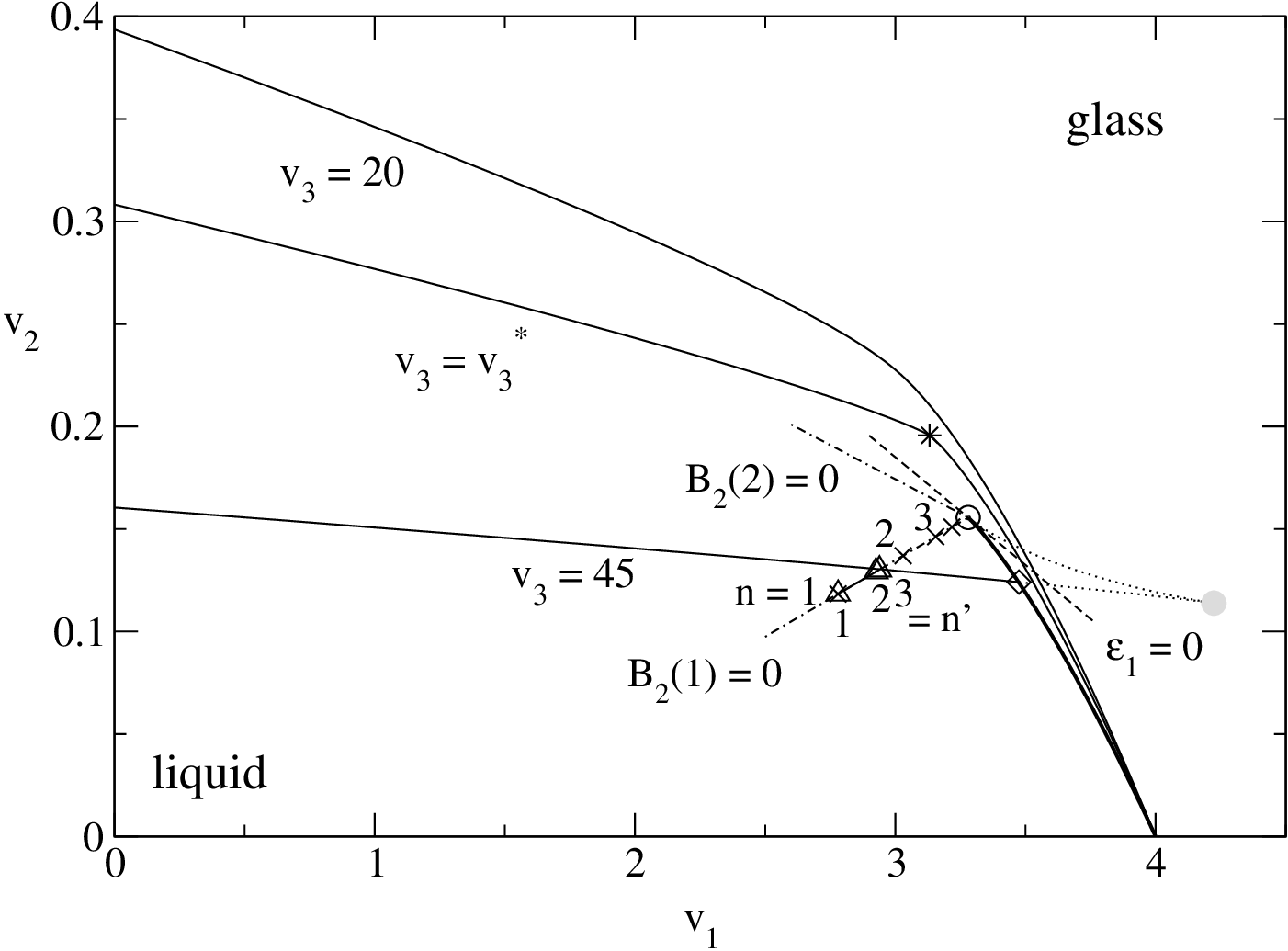}
\caption{\label{fig:BK_PD}Phase diagram for the two-component model defined in 
Sec.~\ref{sec:BK}. The full lines are cuts through the bifurcation surface for
$v_3 = 20$, $v_3 = v_3^*$ and $v_3 = 45$, respectively. 
The value $v_3^* = 24.7\dots$
denotes a coordinate of the $A_4$ singularity indicated by a star. For 
$v_3 > v_3^*$, there occur $A_3$ glass-transition singularities as indicated 
for $v_3 = 45$ by an open circle. The transition lines exhibit a crossing point
shown as open diamond. The dotted lines that join in the cusp
singularity marked by a shaded circle complete the bifurcation diagram of 
Eq.~(\ref{eq:singularity}), but they have no relevance for the discussion of 
the MCT solutions (see text). The dashed line denotes the $v_3=45$ cut through
the surface of vanishing first separation parameter $\epsilon_1$. The 
dashed-dotted lines are the cuts $v_3=45$ through the surfaces of vanishing 
leading correction term $B_2(q) = B_2 + K_q B^2$ in Eq.~(\ref{eq:G1G2q}), 
$q=1,\,2$. The crosses labeled $n=1,\,2,\,\dots$ and triangles labeled 
$n'=1,\,2,\,3$ mark states whose dynamics is 
discussed in Figs.~\ref{fig:BK_quad}, and 
\ref{fig:BK_Chen},~\ref{fig:BK_Chen_Fou}, respectively.
}
\end{figure}

Figure~\ref{fig:BK_PD} exhibits three cuts through the parameter space. The cut
$v_3=20$ is typical for sufficiently small values of $x$. The cut through
the bifurcation surface yields a smooth curve of $A_2$ glass-transition 
singularities. The bifurcation surface for such $v_3^{(2)c}$ deals solely with 
the generic scenario for liquid-glass transitions.

The cut shown for $v_3=45$ is representative for sufficiently large values of 
$x$. In this case, the cubic numerator polynomial in Eq.~(\ref{eq:BK_PD_mu2})
has two zeros $y_1(x) < y_2(x)$ above some $y_0$; they can be evaluated
elementarily \cite{Abramowitz70}. The transition line consists of several 
pieces. The first one, obtained for $1/2\geqslant y > y_2(x)$, is shown in 
Fig.~\ref{fig:BK_PD} as heavy full line. It starts at $v_1^{(2)c} =4,\,
v_2^{(2)c}=0$
and ends at the $A_3$ singularity marked by a circle. The second piece
describes bifurcations with $\mu_2<0$ for $y_2(x)>y>y_1(x)$. It connects the
mentioned $A_3$ singularity with a second $A_3$ singularity of 
Eq.~(\ref{eq:singularity}) that is shown as a shaded circle. 
This piece of the line is shown dotted. 
Decreasing $y$ further, one gets a curve with $\mu_2>0$ that joins the second 
$A_3$ singularity with the point $v_1^{(2)c}=0,\,v_2^{(2)c}=3/(2+x)$. This line
exhibits a crossing point with the first line piece mentioned above, that is
shown as diamond. The part between the second $A_3$ singularity and the 
crossing point is shown dotted, and the final piece is shown as light full 
line. The dotted bifurcation lines and the second $A_3$ singularity are 
excluded from the set of glass-transition singularities because of the maximum 
theorem. These items have been added to the figure merely in order to allow the
reader to recognize the familiar swallow-tail scenario \cite{Arnold86}. The 
crossing point organizes three lines of fold singularities. Between the $A_3$
singularity and the crossing point, there is a line of glass-glass transitions.
The continuation of the line to the boundary of the admissible parameter range
$v_2=0$  deals with liquid-glass transitions. Also the third line between the 
crossing point and the parameter-boundary at $v_1=0$ deals with 
liquid-glass transitions. Both lines are characterized by a discontinuous 
increase of the correlators' long-time limits from zero to the positive 
critical glass-form factors $f_q^{(2)c}>0$.

Decreasing $x$ from large values to smaller ones, the two cusp values $y_1(x)$ 
and $y_2(x)$ approach each other. The corresponding parameter vectors 
$V^c=(v_1^c(x),v_2^c(x),v_3^c(x))$ form curves that approach each other with
decreasing $x$ and join at a certain value $x^*$: $y_1(x^*)=y_2(x^*)=y^*$. 
The pair $(x^*, y^*)$ defines the $A_4$ singularity for the model. The 
parameters for this singularity are obtained, if the derivative of the 
numerator polynomial in Eq.~(\ref{eq:BK_PD_mu2}) is zero for $\mu_2=0$. This 
leads to $(x^*-2)(x^*-4)(x^{*4} - 30 x^{*3} + 136 x^{*2} - 168 x^* + 88) = 0$.
The elementary solution for the zeros of the quartic polynomial
\cite{Abramowitz70} determines the coordinates of the swallow-tail singularity
$x^* =24.779392\dots,\,y^*=0.24266325\dots$~. The cut through the transition 
surface for $v_3 = x^*$ is shown in Fig.~\ref{fig:BK_PD} as pair of light full
lines joining at the $A_4$ singularity which is indicated by a star.

\begin{figure}[Hhtb]
\noindent\includegraphics[width=0.4\textwidth]{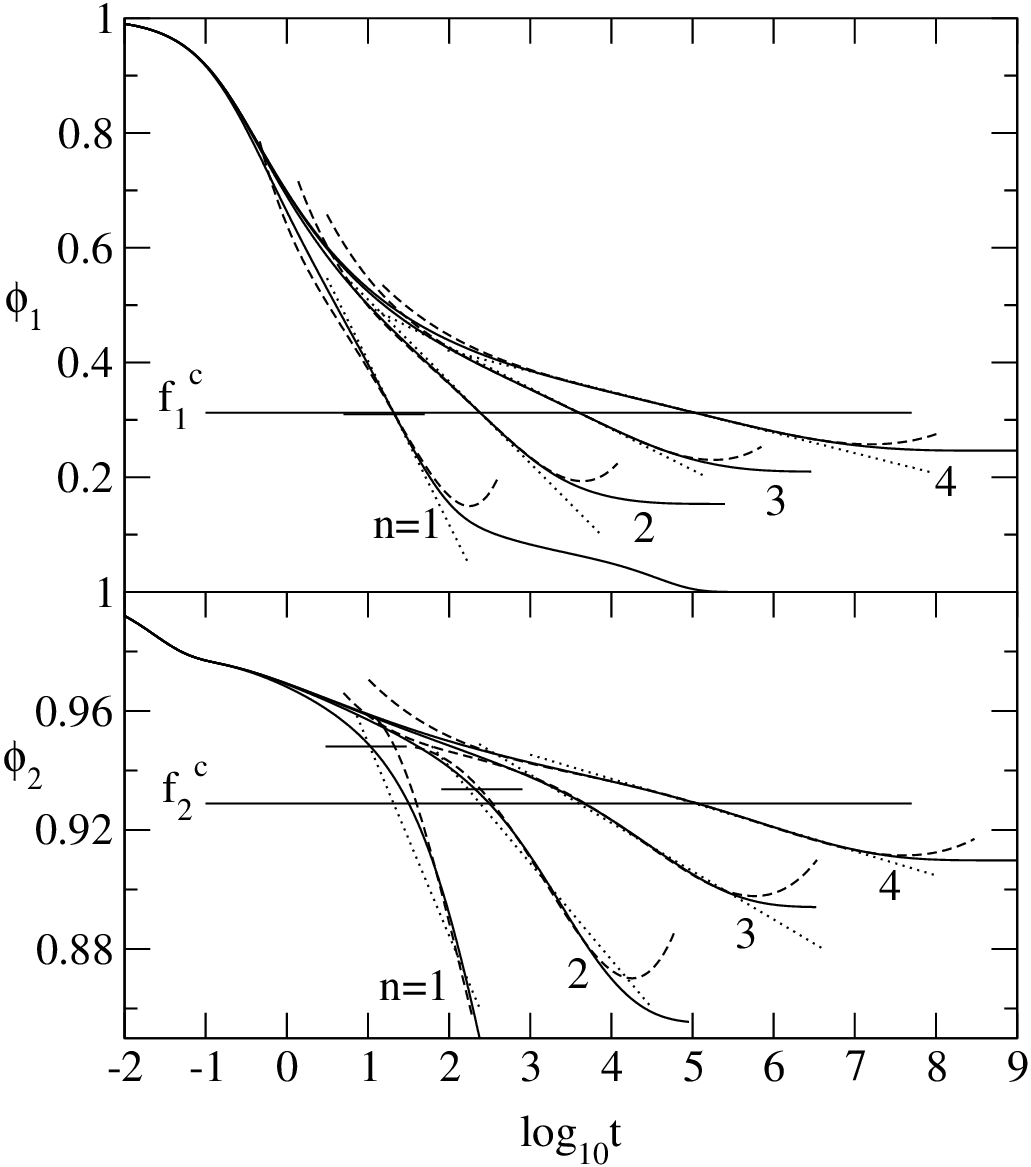}
\caption{\label{fig:BK_quad}Correlators $\phi_{1,2}(t)$ for the two-component
model defined in Sec.~\ref{sec:BK}. The states labeled $n=1,\dots,4$ are 
located on the cut $v_3=45$ through the surface of vanishing dominant 
correction for the first correlator, $B_2(1)=B_2+K_1B^2=0$. The coupling 
constants are $v_1^c-v_1 = 2/4^n,\,v_2^c-v_2 = 0.14907/4^n$ and the states 
for $n=1,\,2,\,$ and $3$ are shown in Fig.~\ref{fig:BK_PD} by crosses.
The full lines are the solution of Eqs.~(\ref{eq:BK_eom}a--c). The dotted
straight lines show the leading approximation, Eq.~(\ref{eq:log_decay}) and the
dashed ones the leading correction, Eq.~(\ref{eq:G1G2q}). The long horizontal
lines show the critical glass-form factors $f_1^c=0.312507,\,
f_2^c=0.92889$ and the short horizontal lines shown for the states $n=1,\,2$
denote the renormalized form factors $f_{1,2}^c+\hat{f}_{1,2}$ according to
Eq.~(\ref{eq:deltafq}). Here and in the following figures, the model is used
with $\tau_1=\tau_2=1$.
}
\end{figure}

Figure~\ref{fig:BK_quad} demonstrates the validity of the factorization 
theorem for states close enough to a cusp singularity $V^c$ and its violation 
for states sufficiently away from it. For the $A_3$ singularity with 
$v_3^c=45$, 
the correction amplitudes calculated from Eq.~(\ref{eq:Kq}) are quite 
different for the two correlators: $K_1=0.06857,\,K_2=-2.049$. Therefore,
the lines for vanishing dominant correction, i.e. the cut of the surfaces
$B_2(q)=B_2+K_q B^2=0,\,\,q=1,\,2$, with the plane $v_3=45$ are quite different
as well, as shown by the dashed-dotted lines in Fig.~\ref{fig:BK_PD}. The four 
states discussed in Fig.~\ref{fig:BK_quad} are chosen on the surface 
$B_2(1)=0$.
Thus, the scenario for the evolution of the $\ln(t/\tau)$ law shown for 
$\phi_1(t)$ is in qualitative agreement with the one discussed in 
Fig.~\ref{fig:A3path_F13}. The states with labels 
$n=3$ and $4$ are so close to the singularity, that the correction term in 
Eq.~(\ref{eq:G1G2q}) proportional to $B_2(2)={\cal O}(\epsilon)$ is not 
important. As a result, the rescaled functions $(\phi_q(t)-f_q^c)/h_q,\,q=1$ 
and $2$ agree for the states $n=3$ and $4$, and the same holds
for the corresponding approximations. However, for the states with labels 
$n=1$ and $2$, the negative coefficient $B_2(2)$ is so large that the 
$\phi_2(t)$-versus-$\log(t)$ curve does not exhibit the straight line obtained 
for $\phi_1(t)$-versus-$\log(t)$ diagram. Rather, the correlator $\phi_2(t)$
exhibits changes of the curvature and inflection points as explained above in
Fig.~\ref{fig:tangentpath_F13} for state $n'=3$.

\begin{figure}[Hhtb]
\noindent\includegraphics[width=0.4\textwidth]{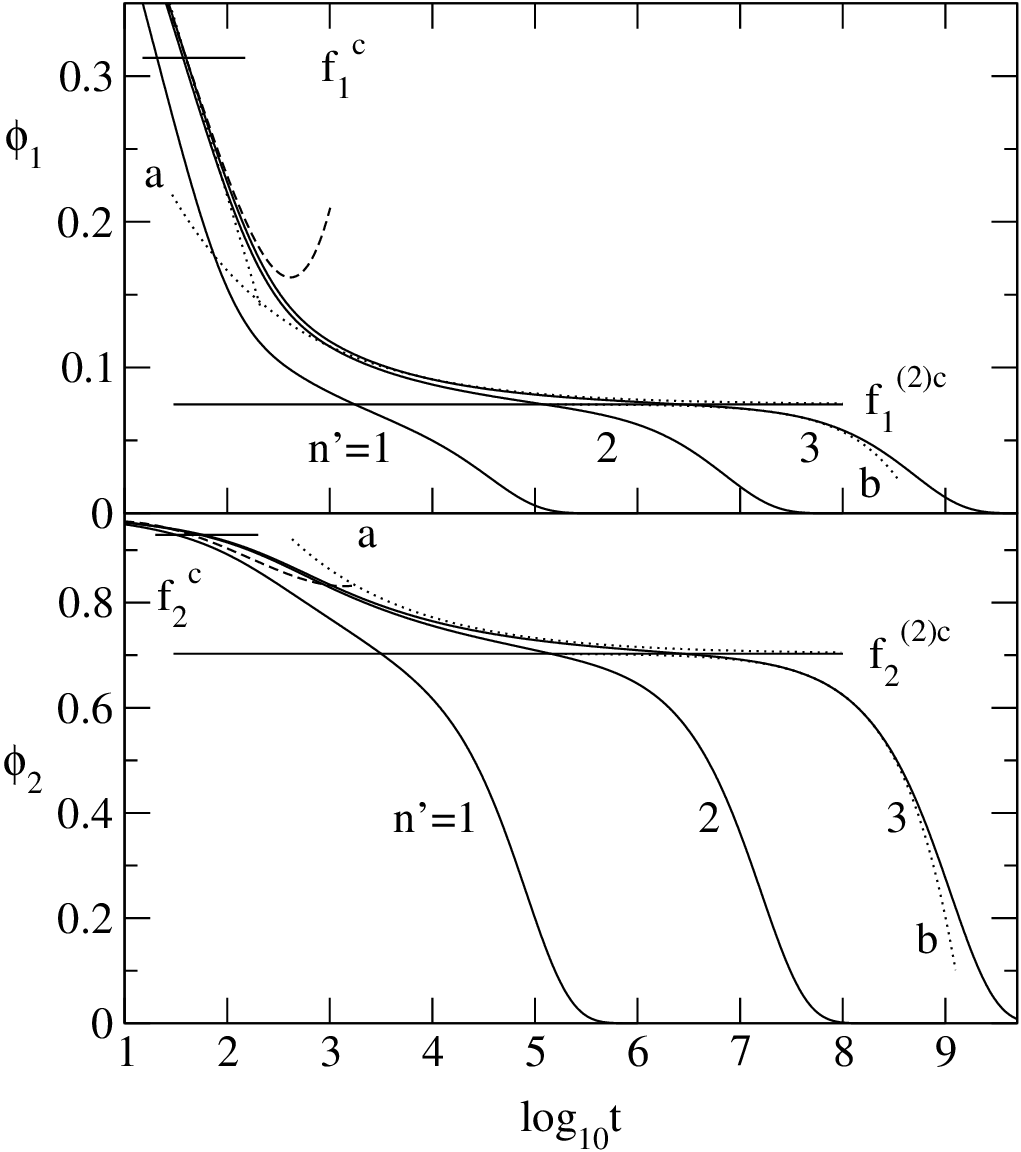}
\caption{\label{fig:BK_Chen}Correlators for the two-component model defined
in Sec.~\ref{sec:BK}. The states with labels $n'=1,\,2,\,$ and $3$ are located
on the line defined by $v_3=45,\,B_2(1)=0$ and have coordinates 
$(v_1,v_2)= (2.7799, 0.1183),\,(2.9254, 0.1292),\,$ and $(2.9391, 0.1302)$,
respectively. They are indicated in Fig.~\ref{fig:BK_PD} by triangles and 
approach the liquid-glass transition point $V^{(2)c}$ with coordinates 
$v_1^{(2)c}=2.941029,\,v_2^{(2)c} = 0.130326$. State $n'=1$ is identical to 
state
$n=1$ discussed in Fig.~\ref{fig:BK_quad}. The horizontal lines show the 
critical glass-form factors $f_q^c$ and $f_q^{(2)c},\,q=1,\,2$, for the cusp 
singularity $V^c$ and the fold singularity $V^{(2)c}$, respectively. The
liquid-glass transition point is connected with an exponent parameter 
$\lambda=0.603$, leading to the exponents $a = 0.363$ and $b = 0.807$. The 
critical decay laws $(\phi_q(t)-f_q^{(2)c}) = h_q^{(2)c}(t_0/t)^a$ are shown as
dotted lines labeled a. The von~Schweidler laws $(\phi_q(t)-f_q^{(2)c})
/h_q^{(2)c}\propto -t^b$ with a time scale fitted for curve $n'=3$ are shown 
as dotted lines with label b. The straight dotted line in the upper panel 
exhibits the leading asymptotic law, Eq.~(\ref{eq:log_decay}) for $\phi_1(t)$ 
and state $n'=3$; the dashed line shows the result of Eq.~(\ref{eq:G1G2q}). The
dashed lines in the lower panel exhibit the leading-correction formulas, 
Eq.~(\ref{eq:G1G2q}), for $\phi_2(t)$ and states $n'=1$ and $3$, respectively.
}
\end{figure}

Figure~\ref{fig:BK_quad} also exemplifies a problem concerning the choice of 
the time scale $\tau$. The complete solution of Eqs.~(\ref{eq:mct_kernel})
and (\ref{eq:mct_laplace_general}) is unique up to the choice of a 
control-parameter independent time scale. The nonlinear coupling of the 
correlators of different index $q$ requires scale universality. However, if a
time scale like $\tau$ is deduced from some approximation to the equation of 
motion, the error of the approximation will result in violations of the scale 
universality for the approximate solutions. Constructing the approximate 
solutions in Fig.~\ref{fig:BK_quad} --- and also in the upper panel of 
Fig.~\ref{fig:BK_Chen}
--- the time $\tau$ was fixed for the leading approximation from $\phi_1(\tau)=
f_1^c$ and for the leading correction from $\phi_1(\tau)=f_1^c+\hat{f}_1$. The 
errors explained lead to offsets for the second correlator: $\phi_2(\tau)\neq 
f_2^c$ and $\phi_2(\tau)\neq f_2^c+\hat{f}_2$, respectively, for the two 
approximations studied. This explains, e.g., that the dashed line for 
$\phi_2(t)$ for the state $n=1$ does not coincide with the full one.
One could also choose $\tau$ differently, e.g. by requesting 
$\phi_2(\tau)=f_2^c+\hat{f}_2$ as done in the lower panel of 
Fig.~\ref{fig:BK_Chen}.

The transition line which is shown in Fig.~\ref{fig:BK_PD} by the light full 
and almost horizontal curve for the cut $v_3=45$ intersects the line 
$B_2(1)=0$ at some glass-transition singularity 
$V^{(2)c}=(2.94\dots,0.130\dots,45.0)$. For states on the
line $B_2(1)=0$ that are close enough to this singularity, one
gets the standard liquid-glass transition scenario, e.g., the evolution of a 
plateau of the $\phi_q(t)$-versus-$\log(t)$ diagram at the critical glass-form
factor $f_q^{(2)c}$ and an $\alpha$-process for the decay below this plateau. 
The universal bifurcation results for an $A_4$ singularity require that the
plateau values are below the critical form factors of the nearby $A_3$
singularity: $f_q^{(2)c} < f_q^c$. For the example under discussion, one gets
$f_1^{(2)c}=0.0747,\,f_2^{(2)c}=0.7027$ and $f_1^c=0.3125,\,f_1^c=0.9289$. The 
precursor of the liquid-glass transition at $V^{(2)c}$ explains the stretched 
tail exhibited in Fig.~\ref{fig:BK_quad} for the decay of $\phi_1(t)$ below 
$0.1$ for the state $n=1$.

To corroborate the discussion of the preceding paragraph, the correlators with
label $n=1$ are reproduced as curves with label $n'=1$ in 
Fig.~\ref{fig:BK_Chen}.
Two further curves with labels $n'=2$ and $3$ are added. They refer to states 
between state $1$ and the transition point $V^{(2)c}$ as noted in 
Fig.~\ref{fig:BK_PD} by triangles. The diagrams for $\phi_1(t)$ for states
$2$ and $3$ exhibit the two-step-relaxation scenario characteristic for
an $A_2$ bifurcation. The decay for $\phi_q(t)<f_q^{(2)c}$ demonstrates the
superposition principle for the $\alpha$-process, and its initial part can be 
described by von~Schweidler's power law. The decay towards the plateaus 
$f_q^{(2)c}$ for $t>1000$ follows the critical law for the $A_2$ singularity 
$V^{(2)c}$. The universal laws for the dynamics near a fold bifurcation imply 
that the correlators follow the asymptote of the critical law, 
$(\phi_q(t)-f_q^{(2)c})/h_q=(t/t_0)^{-a}$, for short times down to about one 
decade above the end of the transient dynamics, i.e. until about $t=10$. In 
particular, for small times, the correlator for state $n'=2$ should approach 
the one for state $n'=3$. However, these features are not exhibited in 
Fig.~\ref{fig:BK_Chen}. Rather, the $t^{-a}$ law
becomes irrelevant for the description of the dynamics below times around 
$10^3$, where the $A_2$ critical curve crosses the curves describing the
logarithmic laws for the $A_3$ singularity. As a result, there appears a window
between the end of the transient and the beginning of the description by the
$A_2$-singularity results where the correlators are described by 
Eq.~(\ref{eq:G1G2q}). This window deals with an increase in time over about two
orders of magnitude. In this window, the logarithmic decay processes destroy 
the manifestation of the $t^{-a}$ law.

The lower panel of Fig.~\ref{fig:BK_Chen} demonstrates a further implication
of $V^c$ dynamics on the precursors of the liquid-glass transition dynamics.
Even though the time scale for the $\alpha$-process for states $n'=1$ or $2$
exceeds the one for the transient by factors $10^4$ and $10^6$, respectively, 
the correlator $\phi_2(t)$ does not exhibit the two-step scenario for these 
states. Rather, there is a large time interval where the approach towards the 
plateau $f_2^{(2)c}$ follows closely the law $(\phi_2(t)-f_2^{(2)c})\propto 
\ln(t/\tau_{\rm eff})$. This is due to cancellation of two effects: The 
asymptotes 
for the $V^c$ dynamics and for the $V^{(2)c}$ dynamics yield a positive 
curvature, while the onset of the $\alpha$-process causes a negative one. The 
resulting nearly linear-$\log(t)$ variation must not be mistaken as the true 
asymptotic logarithmic law given by Eq.~(\ref{eq:log_decay}).

\begin{figure}[Hhtb]
\noindent\includegraphics[width=0.4\textwidth]{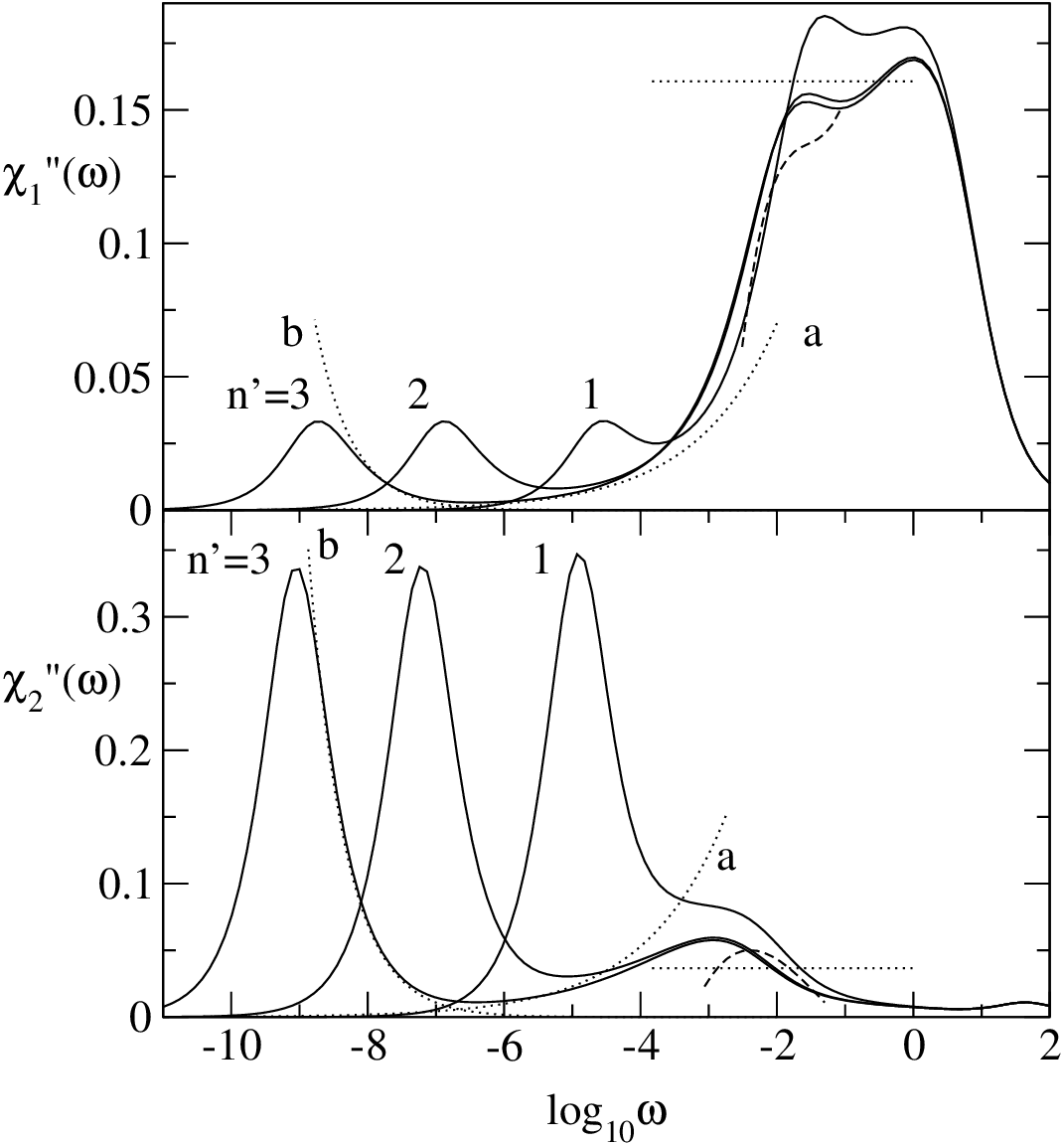}
\caption{\label{fig:BK_Chen_Fou}Susceptibility spectra for the
correlators shown in Fig.~\ref{fig:BK_Chen}.
}
\end{figure}

The destruction of the critical decay law of the liquid-glass-transition 
dynamics by the presence of a higher-order glass-transition singularity 
nearby alters the familiar pattern of the susceptibility spectra, as shown in 
Fig.~\ref{fig:BK_Chen_Fou}. The Debye-peak for the transient dynamics deals 
with the spectra for $\omega > 0.1$, as shown by the peak around 
$\omega\approx 1$ for $\chi_1''(\omega)$. This peak is strongly suppressed and
shifted to higher frequencies for $\chi_2''(\omega)$. There is the large 
frequency regime $-4\leqslant \log_{10}\omega\leqslant -1$, where the 
$\omega^{a}$-law is irrelevant for the description of the structural 
relaxation spectrum. Rather, the critical relaxation
spectrum of the cusp singularity leads to a high spectral enhancement of
$\chi_1''(\omega)$ relative to the $\omega^a$ spectrum; it leads to a second
structural relaxation peak near $\omega\approx 0.01$ in addition to the 
low-frequency $\alpha$-peak. It was discussed in connection with states $n'=3$
and $4$ in Figs.~\ref{fig:tangentpath_F13} and \ref{fig:tangentpath_F13_Fou}, 
that the winding of the 
$\phi(t)$-versus-$\log(t)$ curve around an effective $\ln(t)$-law for states
with $B_2<0$ is a precursor phenomenon of a nearby $A_2$-transition 
singularity. Indeed, the spectrum $\chi_2''(\omega)$ exhibits the $\alpha$-peak
of the mentioned transition with a maximum for $\log\omega\approx -3$. Thus,
because of $B_2(2)<0$, the susceptibility for the second correlator exhibits
two $\alpha$-peaks, referring to the two parts of the liquid-glass-transition
lines discussed in Fig.~\ref{fig:BK_PD}. The low frequency $\alpha$-peaks
shift strongly with changes of $n'=1,\,2,\,$ and $3$, since the states are
shifted towards the transition singularity $V^{(2)c}$ on one of the lines.
The high frequency $\alpha$-peak does not change significantly since the 
distance of the states from the other transition line is almost unaltered.
As explained in connection with Fig.~\ref{fig:tangentpath_F13_Fou}, the leading
correction formula, Eq.~(\ref{eq:G1G2q}), describes the high-frequency wing
of the second $\alpha$-peak.

\section{\label{sec:conclusion}Conclusions}
Describing the states of a system by a vector $V$ of control parameters, the
neighborhood of a glass transition singularity $V^c$ was characterized by a 
sequence of separation parameters $\epsilon_1(V),\,\epsilon_2(V),\,\dots$.
These are smooth functions of $V$ that vanish at $V^c$, and they are considered
as small of order $\epsilon$. The glass-transition singularities $V^c$ are the
bifurcation points of the glass-form factors $f_q(V)=\phi_q(t\rightarrow\infty)
$, i.e. of the long-time limits of the correlators $\phi_q(t)$. These 
bifurcations are of the cuspoid 
family $A_l,\,l=2,3\dots$ . The $\epsilon_1,\,\dots,\,\epsilon_{l-1}$ are the
relevant small coefficients specifying the polynomial of degree $l$ whose 
largest zero determines $f_q(V)$ for small $\epsilon$. The major result
is the proof that the solution of the MCT equations can be asymptotically
expanded in polynomials $P(x)$ of the logarithm of time, $x=\ln(t/\tau)$, for 
states close to an $A_l,\,l\geqslant 3$, and $\epsilon_1(V)<0$. The leading 
term of order $|\epsilon|^{1/2}$ yields the $\ln(t/\tau)$ law, 
Eq.~(\ref{eq:log_decay}). The prefactor for this polynomial of degree $1$ is 
given solely by the first separation parameter $\epsilon_1(V)$. \
The leading correction adds a polynomial of degree $4$. The 
coefficients are of order $|\epsilon|$, and they are determined by 
$\epsilon_1(V)$ and $\epsilon_2(V)$, Eqs.~(\ref{eq:Bcoeffs}) and 
(\ref{eq:G1G2q}). The second correction adds a polynomial of order $7$; the 
coefficients are of order $|\epsilon|^{3/2}$ and determined by $\epsilon_1(V),
\,\epsilon_2(V),\,\epsilon_3(V)$, and so on. Several relaxation scenarios have 
been identified that are utterly different from the MCT scenario for the 
liquid-glass transition. The latter is described by an $A_2$ singularity.

There are distinguished surfaces in parameter space, where the prefactor of the
$x^2$ monomial in the polynomial $P(x)$ vanishes. For this case, the 
$\ln(t/\tau)$ law dominates the dynamics for such times where $\phi_q(t)\approx
f_q(V^c)$. This law may describe the complete decay except for
the transient and for the final exponential approach of $\phi_q(t)$ towards its
long-time limit, Fig.~\ref{fig:A3path_F13}. States near the mentioned
surface exhibit slight deformations of the straight $\phi(t)$-versus-$\log(t)$
curve. There is a concave behavior on one side of the surface and a winding
around the straight line with alternating convex and concave parts on the other
side, as shown for the states $n'=1$--$3$ in Fig.~\ref{fig:tangentpath_F13}. 
The corrections to the leading-order asymptotic results depend on the 
correlator under consideration. The surfaces of dominant $\ln(t/\tau)$ behavior
are different for different correlation functions, as explained in 
connection with Fig.~\ref{fig:BK_quad}.

Every higher-order glass-transition singularity $V^c$ is an endpoint of a 
surface of fold-bifurcation points $V^{(2)c}$ with 
$f_q(V^c)=f_q^c<f_q^{(2)c}=f_q(V^{(2)c})$. For states sufficiently close to 
$V^{(2)c}$, one finds the standard transition scenario with two-step relaxation
described by the interplay of $\alpha$- and $\beta$-scaling laws. The 
correlators for $\phi_q(t)\approx f_q^c$ are a part of the $\alpha$-process. 
Therefore, the logarithmic decay laws as formulated by Eq.~(\ref{eq:G1G2q})
describe the $\alpha$-relaxation master functions. They reduce the range of
validity of von~Schweidler's power law and cause anomalies of the 
$\alpha$-relaxation shape functions as shown for the states $n'=4$--$6$ in
Figs.~\ref{fig:tangentpath_F13} and~\ref{fig:tangentpath_F13_Fou}.

Generically, near a higher-order singularity $V^c$, there is a further surface
of fold bifurcations that crosses the transition surface discussed in
the preceding paragraph, Fig.~\ref{fig:BK_PD}. 
As a result, there is the scenario for transition
singularities of type $A_2$, but now with critical form factors smaller than 
the
ones at $V^c$: $f_q^{(2)c} < f_q^c$. Consequently, the logarithmic decay laws
are a part of the relaxation towards the plateau $f_q^{(2)c}$. They reduce the
range of applicability of the critical decay and introduce a large crossover
interval for structural relaxation between the end of the transient and the 
beginning of the transition dynamics caused by $V^{(2)c}$, as is demonstrated 
in Figs.~\ref{fig:BK_Chen} and \ref{fig:BK_Chen_Fou} for the correlator
$\phi_1(t)$. In particular, there can be a crossover from the transient to a
simple $\ln(t/\tau)$ law followed by a crossover to a von~Schweidler power law.
This scenario was demonstrated by the experiments reported in 
Ref.~\cite{Mallamace00} and by numerical solutions of MCT equations for the 
square-well liquid in Ref.~\cite{Dawson01}. There can also be a susceptibility 
spectrum for structural relaxation consisting of two peaks, as shown for 
$\chi_2''(\omega)$ in Fig.~\ref{fig:BK_Chen_Fou}.

The asymptotic expansion also describes the critical correlator of the
higher-order glass-transition singularity outside the transient,
Fig.~\ref{fig:critical_F13}. These 
correlators deal with the decay towards $f_q^c$ for control parameters at the 
singularity. For states with $\epsilon_1>0$, $\phi_q(t)$ follows the critical
decay until close to its intersection with the long-time asymptote 
$\phi_q(t\rightarrow\infty)=f_q>f_q^c$. Here it crosses over rapidly to the
glass-form factor $f_q$. Summarizing, the formulas of this paper provide a 
qualitative understanding of the decay of the correlations provided the state 
$V$ of the system is close to a higher-order glass-transition singularity
and the correlator $\phi_q(t)$ is close to the glass-form factor $f_q^c$
at this singularity.

Let $L_t$ denote the length of the $\log(t)$ interval where an approximation
by one of the polynomials $P(\ln(t/\tau))$ describes the solution
for the correlator $\phi_q(t)$. Let $L_\omega$ denote the length of the 
$\log(\omega)$ interval where the Fourier-cosine transform of $P(\ln(t/\tau))$
leads to a description of comparable accuracy for the susceptibility spectrum
$\chi_q''(\omega)$. It was explained in connection with 
Fig.~\ref{fig:spectra_A3path_F13} that $L_\omega$ is considerably smaller 
than $L_t$. This phenomenon for glassy dynamics was discussed earlier for the
liquid-glass transition \cite{Franosch97}, but it is more pronounced for the 
higher-order singularities. This feature for stretched relaxation is 
the reason, why it is more difficult to test asymptotic MCT formulas with data
for spectra than it is with data for correlators in the time domain.

If there is a higher-order glass-transition singularity in a disordered system,
there is no generic path for the evolution of the structural relaxation.
Only a parameter surface can be generic for the description of the dynamics 
near a cusp 
singularity. One has to vary two independent parameters to identify an $A_3$
singularity, three parameters to identify an $A_4$ bifurcation, and so on. We 
hope that the demonstration of all basic scenarios for the dynamics near
an $A_3$ singularity will be of use to identify such singularities in colloids,
if there are any. In such case, the derived formulas are elementary enough for
data fitting. Such fitting might lead to a judgment on the relevance of the 
subtle implications of mode-coupling theory for the discussion of 
glass-forming systems.

\acknowledgments
We thank C.~Hagedorn for assistance in preparing the figures and S.-H.~Chong, 
M.Fuchs, A.~M.~Puertas and Th.~Voigtmann for helpful discussions.
Our work was supported by the Deutsche Forschungsgemeinschaft Grant
Go154/13-1.

\appendix
\section {\label{sec:laplace}Laplace Transforms of Logarithms}

The modification of the Laplace transform, introduced in Eq.~(\ref{eq:strafo}),
shall be used to map invertibly functions $F(t)$ of time to functions of the 
complex frequency $z$. The functions are defined for $t > 0$ and ${\rm Im} \,\,
z > 0$, respectively. Euler's second integral for the gamma function 
$\Gamma (y)$ implies ${\cal S} [t^x] (z) = (i/z)^x  \Gamma (1+x)$ if $x > -1$.
Differentiating this identity $n$ times for $x = 0, \, n = 0, 1, 2
\ldots$, one arrives at the formula:
\begin{equation}\label{eq:Strafo_log}
{\cal S}\left [\ln^n (t) \right ] (z) = \sum_k \binom{n}{k}
\Gamma_k \ln^{n-k} (i/z) \,\, .
\end{equation}
Here $\binom{n}{k} = n! / [k! (n-k)!]$ and $\Gamma_k = d^k \Gamma (x=1) / 
dx^k$. One gets in
particular $\Gamma_0 = 1$ and $\Gamma_1 = - \gamma$, where
$\gamma$ is Euler's constant. If $\psi (y)$ denotes the digamma
function, one can write $\Gamma^\prime (y) = \Gamma (y) \psi (y)$.
Iterating this formula, one can express $\Gamma_k$ in terms of the
first $(k-1)$ derivatives of $\psi (y)$ for $y = 1$. The latter
are given by the tabulated values of the zeta function $\zeta (k)$
\cite{Abramowitz70}; for example, $\Gamma_2 - \Gamma_1^2 = \zeta
(2) = \pi^2 /6$. Implications of Eq.~(\ref{eq:Strafo_log}) read
with $n\geqslant 1,n_1\geqslant 1,n_2\geqslant 1$:
\begin{eqnarray}\label{eq:Strafo_log_diff}
{\cal S}\left [\ln^n (t) \right ] (z) - {\cal S}\left [\ln
(t) \right ]^n (z)  = \qquad\qquad\qquad\qquad&& \nonumber\\ 
\frac{\pi^2}{12}\, n (n-1) 
\ln^{n - 2} (\frac{i}{z})  +  \sum_{k=3}^n  \binom{n}{k} \left[ \Gamma_k
- \Gamma_1^k \right] \ln^{n-k} (\frac{i}{z}) \,\, ,&&
\end{eqnarray}
\begin{eqnarray}\label{eq:Strafo_log_power}
&& {\cal S}    \left [\ln^{n_1 + n_2} (t) \right ]  (z) -
{\cal S}\left [\ln^{n_1} (t) \right ]  (z)  {\cal S} \left
[\ln^{n_2} (t) \right ] (z)  = 
\nonumber\\&&\qquad\qquad(\pi^2 / 6) n_1 n_2 \, 
\ln^{n_1+n_2 - 2} (i / z)  \nonumber
\\ & &     + \sum_{k=3}^{n_1 + n_2} \left[ \binom{n_1 + n_2}{k}
\Gamma_k - \sum_l \binom{n_1}{k - l} \binom{n_2}{l}
\Gamma_{k - l} \Gamma_l \right]\nonumber\\
&&\qquad\qquad\qquad\qquad\times \ln^{n_1 + n_2 -k} (i/z)
\,\, .
\end{eqnarray}
These formulas are needed for the evaluation of the function $f^{(2)}(z)$ in 
Eq.~(\ref{eq:f2}).

Specializing Eq.~(\ref{eq:Strafo_log_power}) to $n_1 = n $ and $n_2 = 1$ and 
using the definition of the linear operator  $\cal T$ from 
Eq.~(\ref{eq:Ttrafo}), one gets
\begin{eqnarray}\label{eq:Ttrafo_log_power}
{\cal T}\left [\ln^n (t) \right ] (z)    = && (\pi^2 / 6) \bigl
[ n \ln^{n - 1} (i / z) \nonumber \\  +  \sum_{k=2}^n &&(n - k
+ 1)  \Gamma_{n, k} \, \ln^{(n-k)} (i/z) \bigr ] \,\, ,
\end{eqnarray}
where the coefficients are
\begin{equation}\label{eq:coeff_Gamma_nk}
\Gamma_{n, k} = \binom{n}{k} \bigl [ \Gamma_{ k + 1} -
\Gamma_k \Gamma_1 \bigr ]  / \left [ (\pi^2 / 6) (n - k + 1 )
\right] \,\, .
\end{equation}
Let us construct polynomials $p_n (x)$ of degree $n = 1, 2, \ldots$
obeying Eqs.~(\ref{eq:solution_corr1}). Specializing 
Eq.~(\ref{eq:Ttrafo_log_power}) to $n = 1$ shows that one can choose 
$p_1(x)=x$. Assuming the polynomials for degree $l < n$ to be known, 
Eq.~(\ref{eq:Ttrafo_log_power}) provides the formula for the degree $n$
\begin{equation}\label{eq:pnx}
p_n (x) = x^n - \sum_{k = 2}^n \Gamma_{n, k} \, p_{n + 1 - k}
(x) \,\, .
\end{equation}
Thus, the sequence of $p_n (x)$ can be constructed recursively in
terms of the coefficients $\Gamma_{n, k}$. To derive 
Eqs.~(\ref{eq:Bcoeffs}b--d), one needs:
\begin{subequations}\label{eq:solution_poly}
\begin{eqnarray}\label{eq:solution_poly_p2}
p_2 (x) &=& 2.6160 x + x^2 \,\, ,\\
\label{eq:solution_poly_p3}
p_3 (x) &=& - 2.1482 x + 3.9239 x^2 + x^3 \,\, ,\\
\label{eq:solution_poly_p4}
p_4 (x) &=& - 12.813 x - 4.2964 x^2 + 5.2319 x^3 +  x^4\,\, .
\end{eqnarray}
\end{subequations}

\bibliographystyle{apsrev}

\end{document}